\PassOptionsToPackage{table}{xcolor} %
\documentclass[%
    sigconf,
    authorversion=true, %
]{acmart} 

\newbool{isextendedversion}
\setbool{isextendedversion}{true}
\ifbool{isextendedversion}{\newcommand{\extendedversion}[2]{#1}}{\newcommand{\extendedversion}[2]{#2}}

\usepackage{hyperref}

\usepackage[%
    autopunct=true,
]{csquotes} %

\DeclareQuoteStyle[american]{english} %
        {\itshape\textquotedblleft}
        [\textquotedblleft]
        {\textquotedblright}
        [0.05em]
        {\textquoteleft}
        {\textquoteright}

\SetBlockEnvironment{quote}

\usepackage{tikz}
\newlength\bubblesize
\newcommand\yes{\tikz[baseline=0.1ex] \fill[black] (\bubblesize,\bubblesize) circle (\bubblesize);}

\newcommand\no{\tikz[baseline=0.1ex] \draw[black, line width=0.2ex] (\bubblesize,\bubblesize) circle (\bubblesize-.5\pgflinewidth);}

\usepackage[usetwoside=true]{mdframed}
\usepackage{tcolorbox}
\tcbuselibrary{breakable}
\tcbuselibrary{skins}
\usepackage{enumitem}
\definecolor{darkgray}{gray}{0.3}
\tcbset{}
\newtcolorbox{summaryBox}[2][]
{
    enhanced,
    breakable,
    frame hidden,
    borderline west = {3pt}{0pt}{lightgray},
    colback         = white,
    size            = fbox,
    left            = 0.3em,
    enlarge top by  = 0.3em,%
    coltitle        = black,
    title           = {\color{darkgray} \textbf{#2.} },
    attach title to upper,
    fontupper=\small, 
    #1,
}

\usepackage{makecell}
\usepackage[para, flushleft]{threeparttable}

\usepackage[acronym, %
    nohypertypes={acronym}, %
]{glossaries}
\glsdisablehyper
\newacronym{ai}{AI}{artificial intelligence}
\newacronym{api}{API}{application programming interface}
\newacronym{llm}{LLM}{large language model}
\newacronym{sast}{SAST}{static application security testing}
\newacronym{soc}{SOC}{Security Operations Center}
\newacronym{so}{SO}{Stack Overflow}
\newacronym{sdlc}{SDLC}{software development life cycle}
\newacronym{ssdlc}{SSDLC}{Secure Software Development Lifecycle}
\newacronym{ssd-ses}{SSD-SES}{secure software development self-efficacy score}
\newacronym{irr}{IRR}{inter-rater reliability}

\newacronym{pii}{PII}{personally identifiable information}

\newglossaryentry{copilot}
{
    name=Copilot,
    description={}
}

\newglossaryentry{llama}
{
    name=Llama,
    description={}
}

\newglossaryentry{chatgpt}
{
    name=ChatGPT,
    description={}
}

\copyrightyear{2024}
\acmYear{2024}
\setcopyright{acmlicensed}
\acmConference[CCS '24]{Proceedings of the 2024 ACM SIGSAC Conference on Computer and Communications Security}{October 14--18, 2024}{Salt Lake City, UT, USA}
\acmBooktitle{Proceedings of the 2024 ACM SIGSAC Conference on Computer and Communications Security (CCS '24), October 14--18, 2024, Salt Lake City, UT, USA}
\acmISBN{979-8-4007-0636-3/24/10}
\acmDOI{10.1145/3658644.3690283}

\begin{document}

\title[Using AI Assistants in Software Development: A Qualitative Study on Security Practices and Concerns]{Using AI Assistants in Software Development:\\ A Qualitative Study on Security Practices and Concerns}
\extendedversion{%
    \subtitle{Extended Version}
    \subtitlenote{%
        This is the extended version of the paper originally published at ACM CCS 2024 (DOI: \href{https://doi.org/10.1145/3658644.3690283}{10.1145/3658644.3690283}). It contains additional information and appendices.
    }
}{
    \titlenote{%
        This paper has an extended version: \url{https://arxiv.org/abs/2405.06371}.
    }
}

\author{Jan H.\ Klemmer}
\orcid{0000-0002-6994-7206}
\affiliation{%
  \institution{CISPA Helmholtz Center for Information Security}
  \city{Hanover}
  \country{Germany}
}
\email{jan.klemmer@cispa.de}

\author{Stefan Albert Horstmann}
\orcid{0000-0002-4053-0706}
\affiliation{%
  \institution{Ruhr University Bochum}
  \city{Bochum}
  \country{Germany}
}
\email{stefan-albert.horstmann@rub.de}

\author{Nikhil Patnaik}
\orcid{0000-0002-8055-3750}
\affiliation{%
  \institution{University of Bristol}
  \city{Bristol}
  \country{UK}
}
\email{nikhil.patnaik@bristol.ac.uk}

\author{Cordelia Ludden}
\orcid{0009-0009-8447-5903}
\affiliation{%
  \institution{Tufts University}
  \city{Medford}
  \state{MA}
  \country{USA}
}
\email{cordelia.ludden@tufts.edu}

\author{Cordell Burton Jr.}
\orcid{0009-0000-5596-4518}
\affiliation{%
  \institution{Tufts University}
  \city{Medford}
  \state{MA}
  \country{USA}
}
\email{cordell.burton@tufts.edu}

\author{Carson Powers}
\orcid{0009-0003-6643-8734}
\affiliation{%
  \institution{Tufts University}
  \city{Medford}
  \state{MA}
  \country{USA}
}
\email{carson.powers@tufts.edu}

\author{Fabio Massacci}
\orcid{0000-0002-1091-8486}
\affiliation{%
  \institution{Vrije Universiteit Amsterdam}
  \city{Amsterdam}
  \country{Netherlands}
}
\affiliation{%
  \institution{University of Trento}
  \city{Trento}
  \country{Italy}
}
\email{fabio.massacci@ieee.org}

\author{Akond Rahman}
\orcid{0000-0002-5056-757X}
\affiliation{%
  \institution{Auburn University}
  \city{Auburn}
  \state{AL}
  \country{USA}
}
\email{akond@auburn.edu}

\author{Daniel Votipka}
\orcid{0000-0001-9985-250X}
\affiliation{%
  \institution{Tufts University}
  \city{Medford}
  \state{MA}
  \country{USA}
}
\email{dvotipka@cs.tufts.edu}

\author{Heather Richter Lipford}
\orcid{0000-0002-5261-0148}
\affiliation{%
  \institution{UNC Charlotte}
  \city{Charlotte}
  \state{NC}
  \country{USA}
}
\email{heather.lipford@uncc.edu}

\author{Awais Rashid}
\orcid{0000-0002-0109-1341}
\affiliation{%
  \institution{University of Bristol}
  \city{Bristol}
  \country{UK}
}
\email{awais.rashid@bristol.ac.uk}

\author{Alena Naiakshina}
\orcid{0009-0008-1843-2027}
\affiliation{%
  \institution{Ruhr University Bochum}
  \city{Bochum}
  \country{Germany}
}
\email{alena.naiakshina@rub.de}

\author{Sascha Fahl}
\orcid{0000-0002-5644-3316}
\affiliation{%
  \institution{CISPA Helmholtz Center for Information Security}
  \city{Hanover}
  \country{Germany}
}
\email{sascha.fahl@cispa.de}

\renewcommand{\shortauthors}{Jan H.\ Klemmer et al.}



\begin{abstract}
    Following the recent release of AI assistants, such as OpenAI's ChatGPT and GitHub Copilot, the software industry quickly utilized these tools for software development tasks, e.g., generating code or consulting AI for advice.
While recent research has demonstrated that AI-generated code can contain security issues, how software professionals balance AI assistant usage and security remains unclear.
This paper investigates how software professionals use AI assistants in secure software development, what security implications and considerations arise, and what impact they foresee on security in software development.
We conducted 27~semi-structured interviews with software professionals, including software engineers, team leads, and security testers. 
We also reviewed 190~relevant Reddit posts and comments to gain insights into the current discourse surrounding AI assistants for software development.
Our analysis of the interviews and Reddit posts finds that, despite many security and quality concerns, participants widely use AI assistants for security-critical tasks, e.g., code generation, threat modeling, and vulnerability detection.
Participants' overall mistrust leads to checking AI suggestions in similar ways to human code.
However, they expect improvements and, therefore, a heavier use of AI for security tasks in the future.
We conclude with recommendations for software professionals to critically check AI suggestions, for AI creators to improve suggestion security and capabilities for ethical security tasks, and for academic researchers to consider general-purpose AI in software development. 

\end{abstract}

\maketitle

\section{Introduction}

\Glspl{llm} are among the most notable advances in \gls{ai}. 
\glspl{llm} such as OpenAI's GPT~\cite{chatgpt-blog-intro} or Codex~\cite{OpenAICodex} can generate text and code for given prompts. 
In November 2022, OpenAI introduced \emph{\gls{chatgpt}}~\cite{chatgpt-blog-intro}, a general-purpose AI assistant based on the GPT \glspl{llm} that can also generate code. 
Other tools explicitly target developers, such as \emph{GitHub \gls{copilot}}~\cite{copilot}, which was introduced already in 2021~\cite{copilot-blog-intro}. 
\gls{copilot} is integrated into IDEs to perform automatic completion and generation of code. 
We refer to these \gls{llm}-powered tools as \emph{AI assistants}.

Modern AI assistants are very powerful and can help humans with a few keystrokes, e.g., Copilot was estimated to improve productivity by 30\%~\cite{dohmke2023sea, dohmke2023economic}.
This and the wide availability can also explain the quick adoption by the software industry, organizations, and individual software professionals.
According to the 2023 \gls{so} Developer Survey, about 70\% of professional developers are using or are planning to use AI tools within their development processes and highlight improved productivity and efficiency as main benefits~\cite{2023StackOverflowSurvey}.
Moreover, the survey found developers already use AI assistants in their development workflow, e.g., for writing, testing, debugging, reviewing, or documenting code.

Besides the above benefits, the security performance of \glspl{llm} is overall mixed~\cite{ding2024vulnerabilitydetectioncodelanguage, wu2023exploringlimitschatgptsoftware}.
While research has identified that \glspl{llm} can support security tasks---albeit with various limitations and challenges---such as reverse engineering, CTFs, or other offensive tasks~\cite{shao2024empirical, shao2024NYU, project-naptime, moskal2023llmskilledscriptkiddie, tann2023usinglargelanguagemodels, pearce2022popquizlargelanguage}, AI assistants are also susceptible to generating insecure code~\cite{hajipour2024codelmsec}.
For example, one experiment found \gls{copilot} produced vulnerable code in 40\% of security-critical programming tasks~\cite{pearce2022asleep}, and another showed using AI assistants led participants to produce significantly less secure code~\cite{perry2023ai-assistant-security}.
This is confirmed by other reports~\cite{snyk-ai-2023, conf/usenix/SandovalPNKGD23, majdinasab2023assessing} and known issues, such as AI \emph{package hallucinations}~\cite{package-hallucinations, package-hallucinations-blog} or amplifying insecure codebases by replicating their vulnerabilities~\cite{snyk-copilot-amplifies}.
While the studies mentioned above show that using AI assistants can significantly affect security, they do not explore users' considerations and how they balance security and AI assistant usage.
We argue that those play a crucial role and aim to close this gap with this study.

We further argue that AI assistants can be considered a new source of advice for software professionals. Similar to other advice sources, this might be problematic given that software professionals are known to draw heavily on (online) advice~\cite{Acar:2016ww, acar:2017:secdev, Acar:2017:InternetResources} and that this advice can impact security negatively~\cite{Fischer:2021:ContentReranking, Chen:2019:ICSE:HowReliable}, e.g., when searching for and discussing security issues and solutions~\cite{Yang:2016:WhatSecurityQuestions, Naiakshina:2017} or when copying insecure code snippets from \gls{so}~\cite{Acar:2016ww, Fischer:2017}.
It is unclear how software professionals work with and scrutinize AI suggestions compared to other less-than-perfect sources of advice like \gls{so}.
These concerns are also apparent in the industry.
Google, among other tech giants like Apple~\cite{ai-ban-apple} and Samsung~\cite{ai-ban-samsung}, banned AI assistants, including Google's own \emph{Bard}, from internal usage due to security and quality issues of generated code and privacy concerns~\cite{ai-ban-google}.
As AI assistant suggestions are a new class of advice for software professionals, it is crucial to understand their impact on software security and how they address these security considerations. 

To address these gaps, we conducted a qualitative interview study with 27 software professionals, including software engineers, team leads, and penetration testers, on their experiences with and usage of AI assistants for software development in the context of security.
Additionally, we reviewed the Reddit discussions regarding the use of AI assistants for software development and its potential security impacts. 
We qualitatively analyzed 68~threads and 122~comments relevant to using AI in software development and security practices.
The following research questions direct our study:

\begin{description}\itemsep0em 
    \item[RQ1:] \textit{How are AI assistants used in software development in the context of security?}
    Through interviews with software professionals and reviewing Reddit posts, we investigate how and for which tasks software professionals use AI assistants. 

    \item[RQ2:] \textit{What security concerns and considerations are raised with AI assistants' usage in software development?}
    Our interviews provide insights into the security implications of using AI assistants. 
    We also investigate the role of policy enforcement, code reviewing, and the liabilities associated with insecure code generation.

    \item[RQ3:] \textit{What do developers expect AI assistants' future impact on secure software development will be?}
    Given AI assistants' rapid development and adoption, we asked the participants to speculate about future development and security impact. 
\end{description}

\noindent 
In this paper, we make the following contributions: 

\begin{itemize}[leftmargin=*]
    \item \textit{Qualitative Insights on Security of AI Assistants in Software Development:}
    We are the first to present qualitative insights on how software professionals use AI assistants and consider security in that context. 
    Participants generally mistrust suggestions' security due to overall quality concerns. 
    Nonetheless, they widely consult AI assistants on security-critical tasks (e.g., threat modeling, generating code, vulnerability detection), replacing advice sources like Google and \gls{so}, while critically reviewing suggestions.
    Overall, participants would adopt AI assistants for security tasks if their quality improves.
    The complementing Reddit insights confirm those from the interviews.

    \item \textit{Recommendations:}
    We conclude with recommendations for different AI assistant stakeholders in the context of software development and security.
    In summary, software professionals should remain skeptical and carefully check all AI suggestions, e.g., through peer reviewing and software testing.
    We highlight the need to improve AI suggestion security and recommend AI assistant creators to ensure decent security and reasonable ethical safeguards for security tasks.
    Researchers should focus not only on AI code assistants, but also general-purpose ones and their use in software development.
    
    \item \textit{Artifacts:} For transparency, we provide artifacts for both the interviews and the Reddit analysis (see \nameref{sec:availability} Section).
\end{itemize}

\section{Related Work}
We discuss related work in two key areas:
(i)~security aspects of AI assistants
and
(ii)~AI assistants as a novel advisor for software professionals.
\extendedversion{We provide a detailed background on AI assistants, \glspl{llm}, and their application in software development in \autoref{app:background}.}{}

\subsection{Security of AI Suggestions}

Several studies raise security concerns when using AI assistants, such as for code generation. 
\citeauthor{pearce2022asleep} assessed bugs introduced by GitHub Copilot due to the unvetted code datasets on which the \gls{llm} was trained and found 40\% of generated code to be vulnerable~\cite{pearce2022asleep}. 
The study concluded by advising developers to ``stay awake'' while using the tool as a copilot.
A 2023 replication study found that the proportion of insecure code suggestions decreased from 36.54\% but remains high at 27.25\%~\cite{majdinasab2023assessing}.
Despite these potentially insecure code suggestions, \citeauthor{conf/usenix/SandovalPNKGD23} found in an experiment on writing C code that using AI assistant code suggestions causes only 10\% more security bugs compared to the control group (not using AI assistants)~\cite{conf/usenix/SandovalPNKGD23}.
However, \citeauthor{perry2023ai-assistant-security} found for five programming tasks in three languages (Python, JavaScript, C) that participants using AI assistants produce significantly less secure code while believing to have written more secure code~\cite{perry2023ai-assistant-security}.
So-called \emph{hallucinations} underline that current AI assistants cannot be blindly trusted. 
For example, security researcher \citeauthor{package-hallucinations-blog} found that AI assistants hallucinate software packages that---when registered---are installed by developers and could be used to distribute malicious code~\cite{package-hallucinations, package-hallucinations-blog}.
Moreover, in an industry survey by Snyk among software professionals, 56.4\% reported that insecure suggestions by AI assistants are common~\cite{snyk-ai-2023}.

Given all those security issues in existing AI assistants, the recommendation for developers to ``stay awake'' is highly important~\cite{pearce2022asleep}.
However, the existing studies only show the current shortcomings, and none explore the considerations of software professionals when using AI assistants. 
We argue that human factors need to be understood and considered so that using AI assistants does not weaken security. 
Therefore, we conduct interviews with 27~industry practitioners, and qualitatively complement prior experimental results~\cite{pearce2022asleep, perry2023ai-assistant-security}.
While prior work mainly investigated AI code assistants~\cite{pearce2022asleep, perry2023ai-assistant-security, conf/usenix/SandovalPNKGD23, majdinasab2023assessing}, we also cover general-purpose AI assistants.

\subsection{Security Advice}

We argue that AI assistants are a new source of advice for software professionals, including security advice.
Over the last decade, research examining software developers has found that developers draw heavily on (online) advice~\cite{Acar:2016ww, acar:2017:secdev, Acar:2017:InternetResources, Fischer:2017, Fischer:2019:StackOverflowHelpful, Fischer:2021:ContentReranking, Chen:2019:ICSE:HowReliable, Yang:2016:WhatSecurityQuestions, Naiakshina:2017}.
Researchers found this advice to influence the security of software~\cite{Acar:2017:InternetResources, acar:2017:secdev, Acar:2016ww}.

Software developers discuss security topics on \gls{so}~\cite{Yang:2016:WhatSecurityQuestions}---despite the site containing an almost balanced mix of secure and insecure answers~\cite{Chen:2019:ICSE:HowReliable}.
For example, \citeauthor{Acar:2016ww} found that only 17\% of \gls{so} posts contain secure code snippets and that insecure snippets are copied and deployed in software~\cite{Acar:2016ww}. 
\citeauthor{Fischer:2017} found that insecure code from \gls{so} is widely prevalent in Android apps~\cite{Fischer:2017}.
Besides \gls{so}, \citeauthor{Fischer:2021:ContentReranking} also identified insecure suggestions among top Google search results and demonstrated how re-ranking search results can positively change \gls{so}'s security impact~\cite{Fischer:2021:ContentReranking}.
Particularly interesting in the context of AI assistants, \citeauthor{Fischer:2019:StackOverflowHelpful} demonstrated how deep-learning-based nudging could help developers using \gls{so} to write secure code~\cite{Fischer:2019:StackOverflowHelpful}.

Beyond the security of code, there are other issues with more general security advice that developers can find online. 
In a CCS 2022 keynote, \citeauthor{Mazurek2022} diagnoses an overall security advice ``disaster'' that also affects software professionals~\cite{Mazurek2022}. 
For example, \citeauthor{klemmer2023advice} found usable security advice on the web to be debatable, outdated, or contradicting and might therefore cause insecure implementations~\cite{klemmer2023advice}.
Moreover, researchers found issues in both security advice adoption~\cite{Ion2015, busse_replication_2019} and consensus~\cite{Reeder2017}, and struggles with advice prioritization among software professionals~\cite{redmiles_comprehensive_2020}.

Considering AI assistants as a new advice source that is consulted and directed by humans (e.g., with natural language prompts), the question arises as to whether and how these known challenges of online advice also translate to AI assistants and how this impacts security. 
Our study seeks to answer this question by interviewing software professionals to explore human factors like trust and concerns when using AI assistants for software development.
We argue that understanding such factors is critical as they affect usage behavior and scrutiny when using AI assistants.

\section{Methodology}

This section describes how we designed our study, including our interview recruitment process and line of questioning, our Reddit review process, and our data analysis.

\subsection{Interview Design \& Piloting}

Typical for early exploratory work like ours, we conducted semi-structured interviews, as these enable exploration of key themes but also discussion-led in-depth exploration of novel emerging topics, e.g., by asking follow-up questions and letting participants elaborate their thoughts freely.
We designed an initial interview guide based on our RQs. 
Multiple researchers discussed and revised the interview guide in various iterations to cover all relevant aspects, e.g., adding sub-questions, and enhancing question clarity.
The authors had experience with SE, security, and human factors.
Finally, we validated the interview guide in three pilot interviews with software professionals. 
We included those for analysis, as we did not make any significant changes.

\subsection{Interview Structure}

Below, we outline the structure and content of our interviews. 
The semi-structured interview followed an interview guide split into three sections based on our research questions.
The interview guide is available \extendedversion{in \autoref{app:interview-guide}}{online (cf.\ \nameref{sec:availability} section)}.
We conducted the 27~interviews between July 2023 and March 2024 online via Zoom, lasting an average of 55~minutes (excluding intro and outro). %
Each interview was conducted by one of three interviewing authors. 

\paragraph{Introduction}
At the beginning of each interview, we introduced participants to the interview topic and procedure and obtained consent before recording for later transcription.
We asked them to introduce themselves to get some background information and warm up the participants.

\paragraph{Section 1: Usage of AI Assistants (RQ1)}
First, we asked about AI assistants' use, including the tools the participant had used, their motivation for using them, the tasks for which they used code-AI assistants, and any policies about using AI assistants in their organizations. 
Moreover, we queried participants about their AI assistant workflow, i.e., how they use and approach AI assistants.

\paragraph{Section 2: Security Implications of AI Assistants (RQ2)}
Next, we investigated the participants' understanding, experience, and opinions on the security implications of using AI assistants. 
We prompted participants to discuss security advantages or disadvantages when using AI assistants for software development.
Additionally, we asked about challenges associated with authorship and liabilities, e.g., when an AI assistant would introduce a vulnerability. 

\paragraph{Section 3: Future and Outlook (RQ3)}
In the final third section, we asked participants to elaborate on their outlook on the future of AI assistants and their impact on software development and security. 
We asked how AI assistants have impacted their development process and how they expect this to change. 
We also queried participants about human and AI capabilities by asking whether developers or AI produces more secure code. 
Last, we asked about any needs, desired changes, and wishes for future AI assistants and how they could help with security in software development.

\paragraph{Outro \& Debriefing}
Once the interview was complete, we asked participants if they had any further comments to make and stopped the recording afterward. 
We also asked them to share the study with anyone they know who might be interested. 
After the interview, we sent participants the link to a short, anonymous online demographics questionnaire.

\subsection{Recruitment \& Inclusion Criteria}

To recruit participants, we used our research team's industry connections, hired software professionals on Upwork, and advertised our study at a university, following the recommendations of prior work on developer recruitment best practices~\cite{KaurWhereToRecruit2022}. 
Through Upwork, we advertised to freelancers with experience writing secure code and using AI assistants. 

People who showed interest in the study were directed to a screening questionnaire\footnote{Upworkers were screened directly on Upwork and based on their Upwork profiles. 
We did not screen participants from our professional networks.} that began with the developer screening questions by \citeauthor{dev-screening-questions-time-limits}~\cite{dev-screening-questions-time-limits, dev-screening-questions} and then continued to a series of questions about their current role and experience with software development and AI assistants. 
Participants had to 
(i)~pass two of three random screening questions by \citeauthor{dev-screening-questions-time-limits},
(ii)~be either a developer, team lead, or security expert,
and
(iii)~at least sometimes deal with software security and use AI assistants.
If a participant did not fulfill these criteria, we did not consider them for the interview.
Following the screening, we directed the participants who passed to a consent form explaining the study, outlining the interview's structure, and stating how participant responses would be processed. 
After acknowledging the consent form, the participant was directed to a calendar to select a one-hour slot for an online interview based on their availability and the interviewers' schedule.
We provide both the recruitment materials and the screening questionnaire online (see \nameref{sec:availability} Section).

Each participant was offered compensation in the form of an Amazon voucher worth \$60, or a direct payment via PayPal. Freelancers hired via Upwork were paid \$60 on the platform.

\subsection{Demographics}
We recruited a diverse sample of 27~participants for the interviews: 12 from Upwork, 12 from our professional networks, and three university students also working in industry.
Of those, six identified as tech or team lead, 14 as software developers, and four as machine learning engineers.
Eight participants were security experts working as security engineers, security testers, or penetration testers.
Occasionally, participants held multiple roles.
On average, participants had extensive software industry experience of 14.6~years (md:~12, min:~2, max:~45).
Participants often have to deal with security: eleven indicated their responsibilities include security all the time, while the rest indicated they consider security at least sometimes.
Accordingly, participants overall achieved on average a \gls{ssd-ses}~\cite{SSD-SES} of 55.4~points (md:~60, min:~20, max:~65). 
Compared to \citeauthor{KaurWhereToRecruit2022}'s \gls{ssd-ses} results for developer samples from Upwork (mean: 24.1) and students (mean: 21.9)~\cite{KaurWhereToRecruit2022}, our participants show high confidence in their secure development skills. 
The sample was roughly divided into full-~(11) or part-time~(5) employees and self-employed freelancers~(17) (multiple answers were possible). 
One person was looking for work, and three were students. 
Most participants resided in the US, followed by India, Pakistan, the UK, Brazil, the UAE, Montenegro, Poland, Turkey, and Ukraine.
The majority are highly educated: nine hold a Bachelor's degree, eight a Master's, and two a doctorate. 
One participant currently attends graduate school; the remaining hold a college degree.
A detailed overview of all participants is given in \autoref{tab:participant-overview}.
We observed no differences in participants' answers due to geographic diversity, as participants widely use AI assistants regardless of country~\cite{2023StackOverflowSurvey}.

\subsection{Interview Analysis}

We transcribed the audio recordings using an internal university service and Amberscript~\cite{amberscript}. %
Amberscript initially creates an AI-based transcript before it is corrected by a human transcriber. 
Additionally, we reviewed the transcripts for any transcription errors and corrected them, e.g., field-specific terms or acronyms.
Upon finalizing a transcript, we destroyed the interview's recording. 

To identify common themes in the software professionals' experiences using AI assistants in software development, we adopted the six-step thematic analysis approach~\cite{braun2006using, clarke2015thematic} by Braun and Clarke. 
After familiarizing themselves with the material by conducting the interview and/or reading the transcripts (step~1), three authors analyzed one transcript to develop an initial codebook inductively (step~2). 
After the first transcript, the coders analyzed the transcripts individually so that two coders independently examined each interview. 
After completing the independent transcript coding, both coders merged and reviewed the coding. 
During these sessions, we discussed new codes and disagreements to arrive at a consensus by the end of the meeting. 
We also began categorizing codes into themes based on their commonalities (step~3). 
In this process, the codebook and higher-level themes developed as we refined them in each iteration with the insights from the newly coded interviews (step~4). 
The codebook and themes were reviewed multiple times during the analysis until we reached saturation and a clear definition for each code and theme (step~5).
We report the themes, their codes, and example quotes in \autoref{sec:results} (step~6).
On average, we assigned 84~codes per interview transcript. 
We provide the final codebook in \extendedversion{\autoref{app:codebook}}{the extended version}.

We do not report \gls{irr}~\cite{journal/acm-hci/McDonaldSF19}; \citeauthor{braun-clarke-faqs} advocate not to use \gls{irr} for their reflexive thematic analysis approach~\cite{braun-clarke-faqs, braun-clarke-interview}.
Other researchers support this~\cite{Byrne_2021}.

\subsection{Reddit Discourse Review}

Next, we investigated online discourse on Reddit about using generative AI assistants and their effect on code security. 
We chose to complement the interviews with a review of online discussions to assess whether sentiments described in our relatively small participant sample were reflected in broader discussions on this key forum. 
While this did not provide many additional insights, it supplemented and reinforced  the findings from the interviews.

\subsubsection{Data Collection}

\extendedversion{
    Initially, we conducted a broad gray literature review on the Internet based on four Google searches\extendedversion{(see \autoref{tab:searchterms} in \autoref{app:othertables})}{}.
    We reviewed the top ten results for each search term, examining the general sentiment toward AI assistants and identifying additional terms related to AI-assisted software development. 
    However, most pages did not discuss how AI assistants were specifically used; instead, they gave general opinions on AI use. 
    Reddit was the main place including developer perceptions of AI-generated code. 
    This is also supported by other studies focusing on Reddit as the platform includes in-depth informal SE discussions~\cite{journal/CSCW/LiLDH21, conf/ICWI/HorawalavithanaBLCOI19, conf/ICRE/IqbalKTS21}.
    Similar platforms, like \gls{so}, only included examples of AI use for coding or debugging support.
    Therefore, we turned our attention to Reddit, where posts identified in our initial search were often focused on specific AI assistant use. 
    
}{%
    In an initial exploratory gray literature review based on Google searches, we found Reddit to be the main place to discuss AI assistant usage including developer perceptions of AI-generated code. 
    This is also supported by other studies focusing on Reddit as the platform includes in-depth informal SE discussions~\cite{journal/CSCW/LiLDH21, conf/ICWI/HorawalavithanaBLCOI19, conf/ICRE/IqbalKTS21}.
    Similar platforms, like \gls{so}, only included examples of AI use for coding or debugging support.
    Therefore, we decided to focus on Reddit.
}

We searched r/compsci, r/programming, r/learnprogramming, and r/Technology, the most popular computer science and programming subreddits (i.e., at least one million members). 
We chose these subreddits due to their large membership and active discussions about trending development topics like generative AI. 
We also searched r/ChatGPTCoding, which focuses explicitly on AI-assisted development and potentially yields more specific discussions.
For each subreddit, we repeated our previous Google searches\extendedversion{ (see \autoref{tab:searchterms})}{}, and additionally new terms based on our interview questions, and common terms identified through our initial gray literature search\extendedversion{}{ (see extended version)}.
We reviewed each returned post whether it discussed the usage of AI assistants for coding and, for relevant posts, we identified themes in AI assistant usage (\autoref{sec:redditanalysis}). 
For each relevant post, we collected the top ten comments, which we also reviewed for relevance. 
Next, we calculated term frequency amongst relevant posts and comments. 
We created additional queries from frequent terms in relevant discussions\extendedversion{(see \autoref{tab:searchterms}, Round~2 queries)}{}. 
We then performed a second round of searches and repeated our relevance assessment of all returned posts and comments. 
We used various search terms, to prevent missing security discussions not containing “security.”
In total, our searches yielded 397~posts and 366~comments. 
Of these, 68~posts and 122~comments were relevant.

\subsubsection{Analysis}
\label{sec:redditanalysis}
To determine the collected Reddit posts' and comments' relevance and extract themes, we followed an iterative, open coding approach~\cite{corbin2014basics}. 
First, three authors analyzed 50~documents (from the initial gray literature review) and posts to develop the codebook. 
Then, two authors independently coded posts and comments in groups of 50 using the initial codebook and allowing additional codes to emerge. 
After each round, the coders met, compared codes, resolved disagreements, updated the codebook as necessary, and re-coded any previously coded documents. 
We calculated Krippendorff's alpha ($\alpha$) to measure \gls{irr}~\cite{kripalpha}. 
This process was repeated for four rounds (i.e., 200~documents), until acceptable reliability was reached ($\alpha=0.91$)~\cite{kripalpha}. 
The remaining documents were divided evenly between two researchers and coded by a single researcher.

\subsection{Limitations and Threats to Validity}

\subsubsection{Interviews}
As usual for interview studies, our work has typical limitations that can affect results, such as self-reporting, social desirability, and participation biases. 
For example, participants might not have shared any forbidden AI assistant usage or overreported the extent to which they validate AI-generated code.
While conducting the interviews in English might reduce the number of potential participants and could skew the results, we think this is an acceptable trade-off as English can be considered the primary language in software development. 
We note that the interviews focused on professional software development contexts within companies and larger organizations and might not apply to other scenarios, e.g., hobbyists or open-source developers.
In line with the overall widespread usage of AI assistants among professionals~\cite{2023StackOverflowSurvey}, our sample includes only few participants who do not use AI assistants professionally (e.g., due to company policy), but for private projects. 
Given the unequal distribution, we possibly gained more insights from AI users than non-users.

\subsubsection{Reddit Analysis}
This review has limitations that are common to similar artifact reviews. 
First, our sample is specific to Reddit. This population is likely more active than other developers and may not represent the whole community. 
However, this higher level of engagement offers an upper bound, as these users are also more likely to consider themselves passionate about new technologies like AI assistants~\cite{reddit-stats}. 
Additionally, Redditors' comments are limited in scope and may not provide full context to describe their thoughts and motivations, as this was not the goal of their original post. 
However, this is complemented by in-depth interview insights.
Finally, our searches are a snapshot of the beginning of widespread AI assistant usage and should be considered in context; software professionals' relationships with AI assistants will likely change.

\subsection{Ethics}
Ethical approval for this study was granted by the ethical and institutional review boards (IRB/ERB) of our institutions. %
The research plan and study procedure adhere to (i)~the ethical guidance of the \emph{Menlo Report}~\cite{menlo} and corresponding ACM policies~\cite{acm-policy}, and (ii)~the EU General Data Protection Regulation (GDPR). 
We stored data with \gls{pii} in a secure, self-hosted storage. 
For transcription, we used internal university and GDPR-compliant services.
Besides informing themselves and acknowledging the consent form before the interview, we also introduced participants to our data handling practices, clarified any open questions, and let them know their participation was entirely voluntary. 
They could skip questions or leave the interview at any time.

\section{Results}
\label{sec:results}

\begin{figure}[t]
    \centering
    \includegraphics[width=\linewidth]{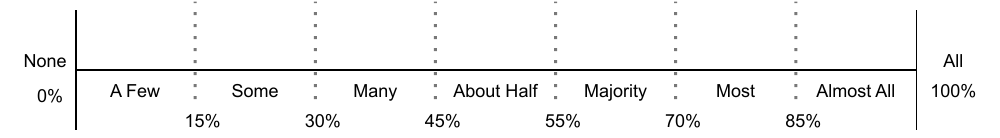}
    \caption{Qualifiers and their respective percentages as used to report our qualitative results. Graphic from \citeauthor{conf/oakland/amft24}~\cite{conf/oakland/amft24}\extendedversion{ as also used by others~\cite{usman23, Zhang_2022} and similarly~\cite{EmamiNaeini2019, Habib2020}}{}.}
    \Description{Qualifier scale used to convey weight for the occurrence of themes and codes in the interviews.}
    \label{fig:qualifiers}
\end{figure}

Below, we detail the results from our qualitative analysis and complement it with additional insights from the Reddit analysis, where appropriate.
For our qualitative interview insights, we do not report exact numbers but rough qualifiers (see \autoref{fig:qualifiers}). 
Exact numbers on individual codes' occurrence can be found in \extendedversion{\autoref{app:codebook}}{our codebook (extended version)}.
For the Reddit analysis, we report descriptive statistics, as the sample was sufficiently large and as we achieved an appropriate \gls{irr}~\cite{journal/acm-hci/McDonaldSF19} (cf.\ \autoref{sec:redditanalysis}).

\subsection{AI Assistant Overview}

To set the general context for the following subsections, we provide an overview of participants' AI assistants.

\subsubsection{AI Assistants}
We found participants widely use AI assistants in their professional work. 
Almost all reported using some AI assistant and doing so very often, most even daily, for various tasks (cf.\ \autoref{sssec:results-tasks}).
Participants mainly use ChatGPT and GitHub Copilot, which aligns with the results of \gls{so}'s 2023 developer survey~\cite{2023StackOverflowSurvey}.
Participants also reported using other general-purpose chatbots by Google (Bard/Gemini) and Microsoft (Bing Chat), but more rarely.
A few participants mentioned other models for coding, such as UniXcoder, Amazon CodeWhisperer, and \gls{llama}. 
However, they were used less often for various reasons, including secondary use of \gls{llama} when dealing with sensitive information or proprietary code that should not be shared with ChatGPT or when the primary AI assistant does not yield the anticipated results. 
Other participants tried assistants (e.g., Amazon CodeWhisperer) for a while, but then abandoned them in favor of ChatGPT or Copilot.
An overview of participants' AI assistants is given in \autoref{tab:participant-overview}.

\subsubsection{Experience with AI Assistants}
All participants reported having used AI assistants previously. 
Most participants started using AI assistants after ChatGPT emerged in late 2022.
Participants became aware of AI assistants through the widespread news and social media coverage around \glspl{llm}; peers, friends, and colleagues using AI assistants; or sometimes, via clients requesting AI features.

\subsubsection{Motivations for Using AI Assistants}
Participants reported several motivations for using AI assistants. 
While some reported security-related motivations, these were rare.
Some participants mentioned that AI assistants could support them as a security expert in their work: 
\blockquote[P1]{For the security point, there are a lot of checks that maybe, as a developer, I couldn't be aware of. Security is exactly one of those points that are not for humans because I believe a lot in machine solutions.}. %
Those participants anticipate AI assistants conducting comprehensive security checks or advising them with more security expertise than they have themselves.

However, most participants were motivated by the increased productivity and time savings when using AI assistants. 
One participant explained this by saying: 
\blockquote[P17]{It can absolutely support us, it can make us more efficient at our jobs, and \textelp{} if I become more efficient, I need less headcount to do the same amount of work.}. %
Along those lines, a few participants reported that using AI assistants can save money---for security, P4 stated that AI assistants save money compared to expensive security scanners. 
Given the productivity improvements, some participants used AI assistants to stay competitive, learn new or enhance their skills in software development.
About half also mentioned a general curiosity in AI and new technology.
Participants also reported certain tasks (discussed in \autoref{sssec:results-tasks}) as use-cases motivating their AI assistant use, e.g., generating code or retrieving information.

\begin{table*}[tp]
    \caption{Overview of the 27~interviews, participants, and AI assistants they use.}
    \label{tab:participant-overview}
    \begin{threeparttable}
        \footnotesize
        \renewcommand{\arraystretch}{0.9}
        \setlength{\tabcolsep}{0.9\tabcolsep}
        \setlength{\defaultaddspace}{0.1\defaultaddspace} %
        \rowcolors{2}{white}{gray!10}
        \centering
        \begin{tabular}{llrllclllllllllllllllllll}
            \toprule
                \rlap{\rotatebox{45}{\textbf{ID}}} & \rlap{\rotatebox{45}{\textbf{Duration}}} & \llap{\rotatebox{45}{\textbf{\#Codes\tnote{1}}}} & \rlap{\rotatebox{45}{\textbf{Recruitment}}} & \rlap{\rotatebox{45}{\textbf{Country}}} & \rlap{\rotatebox{45}{\textbf{Industry Exp.\tnote{2}}}} & \rlap{\rotatebox{45}{\textbf{Role}}} & \rlap{\rotatebox{45}{\textbf{ChatGPT}}} & \rlap{\rotatebox{45}{\textbf{GitHub Copilot}}} & \rlap{\rotatebox{45}{\textbf{Bard/Gemini}}} & \rlap{\rotatebox{45}{\textbf{Self-hosted AI}}} & \rlap{\rotatebox{45}{\textbf{Bing Chat}}} & \rlap{\rotatebox{45}{\textbf{Llama}}} & \rlap{\rotatebox{45}{\textbf{Anthropic Claude}}} & \rlap{\rotatebox{45}{\textbf{Microsoft Copilot}}} & \rlap{\rotatebox{45}{\textbf{CodeWhisperer}}} & \rlap{\rotatebox{45}{\textbf{Tabnine}}} & \rlap{\rotatebox{45}{\textbf{CodeT5}}} & \rlap{\rotatebox{45}{\textbf{UniXcoder}}} & \rlap{\rotatebox{45}{\textbf{Code Bird}}} & \rlap{\rotatebox{45}{\textbf{Colab AI}}} & \rlap{\rotatebox{45}{\textbf{Jarvis}}} & \rlap{\rotatebox{45}{\textbf{Jurassic-1}}} & \rotatebox{45}{\textbf{JupiterOne J1 AI}} \\
            \midrule
                P01 & 00:46:54 & 52 & Network & Italy & >25 & SW Eng. & \yes & \no & \no & \no & \yes & \no & \no & \no & \no & \no & \no & \no & \no & \no & \no & \no & \no \\
                P02 & 00:50:34 & 69 & Network & France &  0--5 & SW Sec. Eng. & \yes & \no & \no & \no & \no & \no & \no & \no & \no & \no & \no & \no & \no & \no & \no & \no & \no \\
                P03 & 00:49:30 & 69 & Network & Viet Nam & 11--15 & Director & \yes & \no & \no & \no & \no & \no & \no & \no & \no & \no & \no & \no & \no & \no & \no & \no & \no \\
                P04 & 00:43:45 & 97 & Upwork & India & 16--20 & Director & \yes & \no & \yes & \yes & \no & \no & \no & \no & \no & \no & \no & \no & \no & \no & \no & \no & \yes \\
                P05 & 01:04:12 & 107 & Upwork & UK &  6--10 & SW Eng. & \yes & \yes & \no & \no & \no & \no & \no & \no & \no & \no & \no & \no & \no & \no & \no & \no & \no \\
                P06 & 00:39:37 & 75 & Student & USA &  0--5 & SW Eng. & \yes & \no & \yes & \no & \no & \no & \no & \no & \no & \no & \no & \no & \no & \no & \no & \no & \no \\
                P07 & 00:53:18 & 97 & Student & India & 6--10 & SW Eng. & \yes & \no & \yes & \no & \no & \no & \no & \no & \no & \no & \no & \no & \no & \no & \no & \no & \no \\
                P08 & 00:54:44 & 81 & Network & UK & 11--15 & Sec. Expert & \yes & \yes & \no & \no & \no & \no & \no & \no & \no & \no & \no & \no & \no & \no & \no & \no & \no \\
                P09 & 00:51:34 & 103 & Student & USA & 6--10 & SW Eng. & \yes & \no & \yes & \yes & \no & \no & \no & \no & \no & \no & \no & \no & \no & \no & \no & \no & \no \\
                P10 & 00:57:31 & 58 & Network & Italy &  6--10 & SW Eng. & \yes & \yes & \no & \no & \no & \yes & \no & \no & \yes & \no & \no & \no & \no & \no & \no & \no & \no \\
                P11 & 00:58:53 & 68 & Network & USA &  >25 & Sec. Expert & \yes & \yes & \no & \no & \no & \no & \no & \no & \no & \no & \no & \no & \no & \no & \no & \no & \no \\
                P12 & 00:58:18 & 127 & Network & Canada & 6--10 & Tech Lead & \yes & \no & \no & \no & \yes & \no & \no & \no & \no & \no & \yes & \yes & \yes & \no & \no & \no & \no \\
                P13 & 00:43:06 & 67 & Upwork & India & 0--5 & Pen. Tester & \yes & \no & \yes & \no & \no & \no & \no & \no & \no & \no & \no & \no & \no & \no & \no & \no & \no \\
                P14 & 01:00:56 & 80 & Upwork & USA & 0--5 & SW Eng. & \yes & \yes & \no & \no & \no & \no & \no & \no & \no & \no & \no & \no & \no & \no & \no & \no & \no \\
                P15 & 00:45:36 & 76 & Network & Brazil & 21--25 & SW Eng. & \yes & \yes & \no & \no & \no & \no & \no & \no & \no & \yes & \no & \no & \no & \no & \no & \no & \no \\
                P16 & 01:06:12 & 129 & Network & UK &  6--10 & SW Eng. & \yes & \yes & \no & \no & \yes & \no & \no & \no & \no & \no & \no & \no & \no & \no & \no & \no & \no \\
                P17 & 00:45:15 & 104 & Network & UK & 16--20 & Director & \yes & \yes & \yes & \yes & \no & \no & \yes & \no & \no & \no & \no & \no & \no & \no & \no & \yes & \no \\
                P18 & 01:10:28 & 46 & Network & USA &  21--25 & SW Eng. & \yes & \yes & \no & \no & \no & \no & \no & \yes & \no & \no & \no & \no & \no & \no & \no & \no & \no \\
                P19 & 01:02:45 & 129 & Upwork & India & 21--25 & Tech Lead & \yes & \yes & \no & \no & \no & \no & \no & \no & \no & \no & \no & \no & \no & \no & \no & \no & \no \\
                P20 & 01:11:30 & 74 & Network & USA & 11--15 & SW Eng. & \yes & \yes & \yes & \yes & \no & \yes & \no & \no & \no & \no & \no & \no & \no & \no & \no & \no & \no \\
                P21 & 01:11:02 & 83 & Upwork & UAE & >25 & Tech Lead & \yes & \yes & \yes & \yes & \yes & \no & \no & \yes & \no & \no & \no & \no & \no & \yes & \yes & \no & \no \\
                P22 & 01:00:54 & 117 & Upwork & USA & 16--20 & Tech Lead & \yes & \no & \no & \yes & \no & \yes & \yes & \no & \no & \no & \no & \no & \no & \no & \no & \no & \no \\
                P23 & 00:45:57 & 58 & Upwork & Montenegro & 0--5 & ML Eng. & \yes & \no & \no & \no & \no & \yes & \no & \no & \no & \no & \no & \no & \no & \no & \no & \no & \no \\
                P24 & 01:07:18 & 85 & Upwork & Pakistan & 6--10 & Sec. Expert & \yes & \no & \no & \no & \yes & \no & \no & \no & \no & \no & \no & \no & \no & \no & \no & \no & \no \\
                P25 & 00:50:05 & 77 & Upwork & Poland & 11--15 & Sec. Expert & \yes & \no & \yes & \no & \no & \no & \no & \no & \no & \no & \no & \no & \no & \no & \no & \no & \no \\
                P26 & 01:12:57 & 102 & Upwork & Turkey & 16--20 & Pen. Tester & \yes & \yes & \no & \no & \no & \no & \no & \no & \no & \no & \no & \no & \no & \no & \no & \no & \no \\
                P27 & 00:25:53 & 51 & Upwork & Ukraine & 0--5 & SW Sec. Eng. & \yes & \no & \no & \no & \no & \no & \no & \no & \no & \no & \no & \no & \no & \no & \no & \no & \no \\
            \midrule
                \rlap{\textbf{Sum}} &  & \llap{2,281} &  &  &  &  & 27 & 13 & 9 & 6 & 5 & 4 & 2 & 2 & 1 & 1 & 1 & 1 & 1 & 1 & 1 & 1 & 1 \\
             \bottomrule
        \end{tabular}
        \begin{tablenotes}
            \item [\yes] Used the respective AI assistant.
            \item [\no] Did not use it. 
            \item [1] Number of codes assigned to the interview transcript.
            \item [2] Software industry experience in years.
        \end{tablenotes}
    \end{threeparttable}%
\end{table*}

\subsection{Usage of AI Assistants}

We asked participants how they used AI assistants for software engineering and security (RQ1).
Below, we report the tasks they perform with AI assistants, participants' associated concerns, how their professional context constrains usage, and how they validate AI-generated code.

\subsubsection{Tasks}
\label{sssec:results-tasks}

Similar to the motivations above, we find---while some participants use AI assistants for security-specific tasks in software development (e.g., threat modeling, identifying vulnerabilities)---participants used AI assistants for many tasks throughout the \gls{sdlc}, e.g., generating code, writing documentation or requirements.
Although the latter are not primarily security-focused, they have security implications.

\textbf{Security Tasks:} 
Many participants reported using AI assistants for security tasks.
Identifying vulnerabilities and fixing security bugs in code were mentioned the most by some participants: 
\blockquote[P15]{If you just grab some pieces of code that are exploitable and just paste them in ChatGPT \textelp{} Probably the AI is going to find out some changes for you to make the code more secure.}. %
Some participants used AI assistants earlier in the \gls{sdlc}. P4 mentioned using ChatGPT for threat modeling:
\blockquote[P4]{What we are doing is using all this prompt engineering and giving as much information as possible to the tool and help us in defining the threat models rather than doing manual work.}. %
While P4 mentioned AI assistants are not perfect, they at least provide a solid starting point for manual refinement, and sometimes, they would have forgotten the AI-suggested attack vectors otherwise. P15 also mentioned using AI assistants to create an exploit, and P12 mentioned AI assistants helped them explain results from static analysis tools.
We found our software engineers used AI assistants slightly more for security tasks, like checking for vulnerabilities, than those primarily focused on security, e.g., security testers.
One explanation is that participants with a strong security background assume AI performs worse on security tasks.

Notably, only two Reddit posts specifically targeted security and no comments. 
While rarely discussing security explicitly, commenters pointed out that AI-generated code is often of low quality and should not be relied on (P=3, C=24)\footnote{\textit{C} denotes the number of comments about a topic, \textit{P} the number of posts.}, which could include security issues. 
One commenter explained \enquote{It \textins{AI} will lead to tech debt and shabbily maintained and written code.} 
This AI skepticism was the most common response to posts indicating a use or interest in using AI assistants for code generation. 
On average, posts about code generation received 0.62~comments indicating AI-generated code should be thoroughly scrutinized.

\textbf{Coding-Related Tasks:} 
Most often, almost all participants reported using AI assistants for coding-related tasks.
Almost all participants used AI assistants to generate code, as in this example:
\blockquote[P12]{In most cases, I found the code written by ChatGPT to be a very good starting point. It's never perfect, but a very good starting point that we just need to add a few if-else statements to catch some edge cases or to fill in our credentials for certain databases.}. %
This sentiment was common on Reddit, as 33~posts described using AI assistants to generate code or wanting to learn how to use AI assistants. 
Similarly, this was the second-most common topic of AI assistant-related Reddit comments (C=30). 
The only more common comment topic was whether AI assistants would replace developers altogether (P=9, C=48).
Regarding code generation, a few participants stated that they used AI assistants to translate code into other programming languages. 
At the same time, participants said they might not understand code for unfamiliar programming languages:
\blockquote[P2]{I have close to no experience with Rust and I guess I can ask ChatGPT to produce Rust code for me, even though I would not be able to actually understand whether it's correct or not.}. 

Some participants reported using ChatGPT for debugging code, fixing bugs, or explaining code:
\blockquote[P17]{The thing I love about things like ChatGPT is it doesn't just give you the answer, it explains why, especially if you're asking it for code it will tell you: \textelp{} here's what it does.}. %
Minor other use cases were related to code quality, such as refactoring, optimizing, and reviewing code.
These uses were reflected in the Reddit discourse, with several posters and commenters describing using or wanting to use AI assistants for debugging (P=7, C=0) or explanations (P=7, C=4). Few discussed refactoring (P=2, C=0) and optimizing (P=4, C=10). 
One Redditor preferred using ChatGPT to do more straightforward coding tasks, saying they
\blockquote[redditor]{\textins{Use} ChatGPT for automating the boring stuff like code refactoring, unit tests, and code documentation}.

\textbf{Information \& Advice Source:} 
The majority of participants reported consulting AI assistants as general sources for researching information, asking questions, and obtaining advice.
About half of the participants said they use AI assistants as replacements for search engines (like Google) and online communities (like \gls{so}). 
Two participants fittingly describe it:
\blockquote[P17]{Wherever I would formerly use StackOverflow, I now use OpenAI \textins{ChatGPT}.} %
and 
\blockquote[P12]{Previously \textelp{} you would say, \enquote{Have you Googled it?} Nowadays we'll say, \enquote{Did you ask ChatGPT?}}. %

\textbf{Documentation \& Requirement Analysis:} 
Lastly, we found the majority of participants used AI assistants to perform tasks supporting application design and development in the \gls{sdlc}---beyond coding. 
This includes requirement analysis, creating documentation and reports, and writing Jira stories (e.g., for bugs/issues).
A few participants who conducted security tests said they write their security reports with AI assistance.
Facing those tasks likely explains the higher popularity of ChatGPT compared to GitHub Copilot.

\textbf{Tasks AI Assistants are not Used for:} 
Some participants explicitly said not to use AI assistants for the above-mentioned tasks. 
For example, some participants stated not to use it for discovering vulnerabilities: 
\blockquote[P13]{Vulnerability wise, I don't think it does that good, just to identify that source code wise. \textelp{} Other premium scanners or some things would be doing a better job, I guess.}. %
The concerns about AI assistant performance and capabilities were also prevalent among other participants and prevented them from AI-assisted bug fixing, code reviews, or threat modeling: 
\blockquote[P25]{I even tried to use for the threat modeling, but it was so bad that it was \textelp{} just a nightmare. I just \textelp{did} it from scratch by myself.}. %

\textbf{Security Relevance of Non-Security Tasks:}
Considering code generation, the primary task is not security, but using insecure AI-suggested code might have severe security consequences. 
However, some participants explicitly mentioned that generated code has almost no security impact, as they would only create smaller snippets or not use them in production.
P15 explained: \blockquote[P15]{I don't generate one page of code. It's just a few lines of code \textelp{}. Those are small functions; they don't have security concerns at all.}. %
Participants predominantly expressed that security issues would be easier to spot when generating smaller code chunks, which is their typical use case. %
We cannot assess this hypothesis without future research---the participants' experience might be correct. 
Still, small snippets might be dangerous, e.g., when AI hallucinates packages~\cite{package-hallucinations, package-hallucinations-blog}.
Besides code generation, searching and looking up information could be security-relevant; if the AI output is incorrect, software professionals might make decisions that undermine security.

\subsubsection{Organizational Context \& Privacy Constrain AI Assistant Usage}
\label{sssec:results-professional-context}

While most participants use AI assistants daily for various tasks, they reported that the professional context in their organization can constrain how AI assistants are used. 
When asked about security, participants did not mention constraints due to the security of the software they create but mainly privacy, legal, and indirect security concerns when using third-party AI assistants.

The primary concern among most participants was leaking sensitive data when using third-party AI assistants---either that the AI provider is breached or their inputs are used for training \glspl{llm} and might be reproduced by future models.
A few participants stated that they feared their code or internal knowledge might be leaked and used by attackers, e.g., to find and exploit vulnerabilities in their code. %
Due to these concerns, many participants reported using AI assistants only in one direction: using the AI-generated code, but never supplying their code to the model.

However, participants were mainly concerned about leaking proprietary information and code, confidential company data, violating non-disclosure agreements, license agreements, or other contracts, or leaking otherwise protected data or \gls{pii}.
Consequently, participants police themselves on what they supply as inputs to AI assistants.
P09 describes this fittingly: 
\blockquote[P09]{I cannot simply copy code to a ChatGPT or other AI assistant because they're going to put it on a larger pool of data and supply it everywhere. The security measures in our company or any company in general wouldn't permit us to do those things. We'll have to break down the problem statement and essentially ask only the context, as if asking another person who is not in our company.}. %
Due to leakage concerns, some organizations set AI usage policies (\autoref{sssec:policy-regulations}) or desire self-hosting models or getting privacy guarantees (\autoref{sssec:results-wishes}).
While this sentiment was not common on Reddit, one commenter, in response to a post about using AI assistants, warned 
\enquote{%
\textins{it} returns entire snippets of copyrighted code without any attribution.} 
We expect the discrepancy between interviews and Reddit comes from the focus on internal organization policies.

Participants reported other minor constraints that were largely unrelated to security. 
This includes copyright infringement and intellectual property violations when using AI assistants that reproduce training data and high AI assistant costs, e.g., for subscriptions or operation costs if self-hosting. 
For most, the costs were outweighed by productivity improvements.

\subsubsection{Quality Assurance of AI-Generated Code}
\label{ssec:results-checking-correctness}

Generally, participants reported checking AI-generated outputs, especially code, before using them. 
This is grounded in a general mistrust due to correctness and reliability issues that participants experienced, which they also translate to security (\autoref{sssec:results-concerns}).
One participant said:
\blockquote[P14]{I just think there are vulnerabilities and there are things that it doesn't know. Currently, and at least for the next five years, I think all code written by AI will need to be gone over by a professional.}. %
Overall, we found participants to commonly follow a three-step process to check AI-generated code, as depicted in \autoref{fig:checking-ai-generated-code}: 

\begin{figure*}
    \centering
    \includegraphics[width=0.75\linewidth]{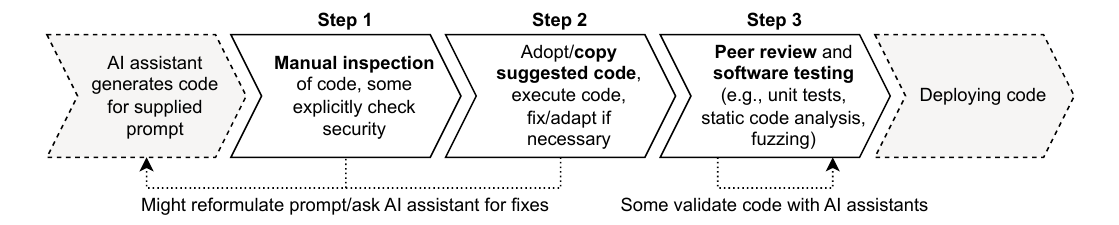}
    \caption{Our participants described three steps to inspect AI-generated code.}
    \Description{Common three-step process that participants apply to check and inspect AI-generated code.}
    \label{fig:checking-ai-generated-code}
\end{figure*}

\textbf{(1)~Manual Inspection: }
First, almost all participants said to inspect the generated code and check for anomalies or issues:
\blockquote[P19]{I don't completely trust them, but I at least read over their code.}. %
While some participants mentioned specifically checking for security issues, the majority was more concerned about functional correctness---one even said not to check security at all.

\textbf{(2)~Copy, Execute, and Fix Suggested Code:} 
Next, many participants reported copying and executing the suggested code to check whether it works.
If not, they fix it manually or with the AI assistant's help. 
However, a few participants indicated not adopting the generated code but using it as a blueprint, re-implementing the final code entirely on their own:
\blockquote[P27]{\textins{I} reuse it without copying, but just reading, understanding how it works, and doing it by myself.}. %

\textbf{(3)~Peer Review \& Software Testing:}
Third, about half of the participants reported that they complement their checks with peer reviews before merging code. 
The respective reviewers varied depending on the organizational structure and resources:
Participants reported other team members, team leads, quality assurance teams, or dedicated security experts/teams.
We suspect a higher prevalence of peer reviewing, as it is a common practice not specific to AI-generated code that participants might not report. 
Some participants said they did not distinguish human and AI-generated code and apply the same reviewing and testing procedures:
\blockquote[P17]{This isn't something that we introduced because of generative AI \textelp{}. We've always had a full SDLC. \textelp{} the same rules apply as every other piece of code you write. Code that's generated by the AI goes to the exact same review process \textelp{}}. %

Similarly, many participants used various forms of software testing to validate AI-generated code.
This included classical forms of software testing like unit tests, static analysis tools, and fuzzing.
Some participants asked the AI assistant who generated the code to check it or cross-check it with another AI assistant.

\begin{summaryBox}{Key Findings: Usage of AI Assistants (RQ1)}
    \begin{itemize}[leftmargin=*]
        \item AI assistants are used for various security tasks, such as threat modeling or vulnerability detection, and security-relevant tasks (e.g., code generation) in the \gls{sdlc}. Moreover, participants consult AI assistants with general questions and for advice, replacing \gls{so} and Google.
        
        \item In the corporate context, the main concern is privacy---not security of software created with its help---which constrains AI assistant usage.
        
        \item While AI-generated code is often directly copied to codebases, it undergoes quality assurance similar to human-written code, including peer review and software testing.
    \end{itemize}
\end{summaryBox}

\subsection{Security Concerns \& Considerations}

Below, we cover participants' security concerns and considerations (RQ2).
As participants largely showed mistrust in AI suggestions, we report on reasons for their mistrust. 
Moreover, we report on AI assistant usage policies in the participants' corporate context.

\subsubsection{Security Concerns and AI Assistant Challenges}
\label{sssec:results-concerns}

Overall, participants generally mistrusted AI assistants for security due to several challenges they experienced.

\textbf{Mistrust and Blind Trust:}
Most participants mistrusted AI assistants and their generated suggestions, as they doubt the security of its suggestions or more indirectly the correctness of suggestions: 
\blockquote[P02]{Right now, I will not be able to trust the code produced by ChatGPT in security-sensitive scenarios \textelp{} It's mostly like I cannot really trust the correctness of this code, I'm not sure I should trust the security properties of it either.}. %
Despite this general mistrust, many participants feared negative security impacts as they expect some software professionals might trust AI blindly, not questioning security:
\blockquote[P03]{%
I worry that people will lean on or depend too much on the things generated by AI and forget about the security.}. %
Relatedly, some participants complained that AI assistants are always confident, even if the suggestions are wrong or insecure. 
Instead, participants desired AI assistants to indicate their confidence, similar to how humans would express uncertainty about a solution.
Many participants also mentioned that wrong AI suggestions are hard to recognize.

\textbf{Poor AI Suggestion Quality:}
Participants expressed the above mistrust concerning poor AI suggestion quality---for general suggestions and AI-generated code.
Most participants reported overall quality issues, such as inaccurate, outdated results---due to older datasets on which \glspl{llm} were trained---or hallucinations. 
A few also expressed that quality degraded over time.

Almost all participants expressed quality problems when generating code with AI assistants.
This mainly concerned the necessity to review  and rework code (\autoref{ssec:results-checking-correctness}) as the majority experienced that code did not work as intended or were even concerned that AI might otherwise introduce bugs:
\blockquote[P17]{I don't think we've ever taken anything directly from the AI \textelp{} straight into code. Even after all the checks, I think everything's had to go through and be subtly changed.}. %
Especially for more complex problems and when generating larger amounts of code, about half the participants reported lower quality.
A few participants found AI-generated code challenging to refactor. 
This aligns with some participants who said AI-generated code was hard to understand and fix.

These issues likely relate to the interaction challenges the participants mentioned with AI assistants.
For most participants, that concerned prompting and prompt engineering, i.e., steering the AI assistants to generate what the user desires. 
Many participants reported needing to change their prompts in multiple iterations until they were satisfied with the AI-generated answers.
One participant said:
\blockquote[P08]{You end up just having to both tweak the prompts and then tweak the code \textelp{} to actually fit your purpose.}. %
Fittingly, many participants struggled to provide the \gls{llm} with enough context to create high quality results, mentioning the limited context windows and numbers of tokens that \glspl{llm} can process, or request limits in AI assistants.
P06 summarizes all this well: 
\blockquote[P06]{Code generated by AI \textins{might not} be safe because it does not know the entire code, it just gives relation to my question of what I asked. I don't feel it's completely safe. I do check back.}. %

\textbf{Actual Security Issues through AI Assistants:}
Despite the widespread security and quality concerns, participants rarely reported facing security issues using AI assistants.
Instead, experiences were mixed.
While a few were unsure, some participants did not perceive any change in their projects' security since adopting AI assistants.
Only a few mentioned security improvements, like identifying a new attack vector in their software.
A few others, however, reported security shortcomings in the generated code: 
\blockquote[P08]{For example, it doesn't hash the password, it doesn't add salt to the password unless you specifically tell it to do that. \textelp{} Generally, it's not very secure. We will generally find mistakes in the majority of code snippets that it creates.} %

Nonetheless, others did not experience security issues despite all other quality shortcomings:
\blockquote[P17]{%
It definitely writes bad or suboptimal code in some places, but I don't frequently see glaringly obvious security vulnerabilities being created by the AI.}. %

\textbf{Security Concerns:}
Most participants expressed security concerns when using AI assistants. 
Many doubted its ability for security tasks, as one said: \blockquote[P26]{I would never rely on ChatGPT for security.}. %

First, about half the participants found that AI assistants will likely introduce security issues through generated code:
\blockquote[P02]{I think the more code will be produced by the current existing AI tooling, the more we'll see security bugs in them, and we'll probably see new patterns of security bugs.}. %

Second, some participants were concerned about \gls{llm} poisoning and facing models trained to create insecure suggestions. 
A few outlined that they would not be able to recognize this:
\blockquote[P04]{What if it is developed using a data set that is inherently vulnerable? \textelp{} Those challenges are there.}. %
Similarly, a few were concerned about AI assistants acting as a malicious dependency:
\blockquote[P08]{%
Actually, anything that's produced also has that vulnerability.}. %

Third, some participants questioned AI assistants' suitability for security at all or partly:
\blockquote[P24]{ChatGPT is much more limited in cyber security, but it's very good in code generation and programming.}. %
For example, participants found it performs worse than \gls{sast} tools in finding vulnerabilities; P13 and P26 argued that current AI models, like GPT and Codex, cannot handle more complex security tasks. 
A few participants experienced that ChatGPT needs to be actively directed into a \enquote{security mindset} in their prompts and questioned why this is not a default:
\blockquote[P27]{Honestly, I'm not so satisfied with the security level of the code because ChatGPT doesn't include it by default. If you don't ask it, it doesn't include the security steps in the code.}. %

\subsubsection{AI Assistants' Ethics Safeguards on Security Tasks:}
\label{sssec:results-ethics-constraints}
Some participants reported that the \emph{ethical safeguards} built into AI assistants (also called \emph{guardrails} or \emph{constraints}) rejected their prompts for security tasks, e.g., finding vulnerabilities.
However, several participants explained that they had reframed their prompts to circumvent the AI assistants' safeguards and have it assist with a vulnerability:
\blockquote[P07]{In ChatGPT, for example, if I'm asking how to do an SQL injection, it will say \enquote{SQL injection is an unethical thing, sorry, I can't help you with that.} If I'm asking in some \textelp{} indirect way, it will explain to me all the details.}.
Therefore, participants perceived it more like a circumventable usability obstacle, but not an actual constraint.
However, this might not be enough to leverage AI assistants for more offensive security tasks, even when circumventing ethics constraints.
For example, P25 explained that AI assistants can suggest possible attack vectors but will not perform the attack for them:
\blockquote[P25]{It gives you some tricks. \textelp{and} possible attack vectors, but it will not make the attack instead of you. If you don't \textelp{} understand how the basic attack works, just using one command that ChatGPT gives you, will not make you a successful attacker or hacker.}.

\subsubsection{Policy \& Regulations on AI Usage}
\label{sssec:policy-regulations}

As large tech companies, such as Apple or Google, have banned the use of AI assistants for security and quality reasons, we asked participants about policies on using AI in their work.

About half of the participants stated their company did not have a policy regulating the use of AI assistants. 
Participants who work in companies that recently started using AI assistants argued that policies were not considered as they needed to test the boundaries and use cases of AI assistants. 
One participant declines policies and explicitly said: \blockquote[P10]{No, absolutely no. No policy interference!}. 
Self-employed participants did not require a policy, since they work for themselves. %
Some participants advocate for self-policing and argue that a policy is not required when following common sense, e.g., not sharing sensitive data.
A few participants expressed that the need for policies when using AI assistants is security-relevant, especially when using it to generate code that, therefore, needs to be tested (see \autoref{ssec:results-checking-correctness}).
These participants desire a standardized verification process to evaluate and verify the security of the generated code before merging it to the main code base. 

Besides policy absence, other participants reported having policies.
A few even reported that their companies banned AI assistants by policy, but mainly for privacy reasons, as outlined in \autoref{sssec:results-professional-context}.
One explained how their company developed a custom AI assistant based on public \gls{llm} APIs, which serves as a proxy that filters requests based on the company policy before sending the prompt to the third-party \gls{llm}.
Another participant explained that AI assistants remain forbidden by default but can be used when clients authorize their data to be shared with AI assistants. 

Participants rarely reported shadow practices, while a few assumed AI assistants usage anyway, even if forbidden.
P16 justified their use of ChatGPT, even though their company requires Bing Enterprise because they perceived the former to perform better.

\subsubsection{Responsibility for AI-Generated Code}
For a fictive scenario, we asked about responsibility when AI-generated code is used in production, only to find it vulnerable and exploited later. 
Almost all participants agreed the human who uses the AI assistant remains responsible, while a few said their company is also responsible. 
Participants mainly argued that an AI assistant is a tool and does not replace the developers' agency, but developers must check suggestions before using them.
P01 compared it to copying and pasting code from \gls{so}, stating that it is the human's responsibility that the code works as intended. 
P08, however, argued that their company would be responsible as it should have ensured a code review process, specifically for AI-generated code.
Only a few other participants said they consider the AI assistant creators responsible.

\subsubsection{Security Performance: Human vs.\ AI}
We asked participants who wrote the more secure code, comparing humans and AI.
Overall, participants' opinions differed.
About half of them argued that humans would create more secure software.
On the contrary, some expected AI to perform better. 
A few participants said that AI assistants currently perform at the level of junior human developers but expect AI to become better than humans. 
While P02 currently expected neither to perform well, some participants perceived AI assistants and humans to complement each other, therefore achieving the best security when AI assists humans:
\blockquote[P14]{I think that the most secure code would be a combination of the two. I think that both I and the model alone would generate code with insecurity. I think that the combination of both me and the model would write the most secure code.}. %

\begin{summaryBox}{Key Findings: Security Concerns \& Considerations (RQ2)}
    \begin{itemize}[leftmargin=*]
        \item When using AI assistants, participants consider security, indicated through many security concerns, but only a few faced actual security issues. 
        The overall (security) mistrust in AI assistants is primarily due to code quality concerns.

        \item Participants mainly assume humans to perform better security-wise than AI assistants, and perceive humans to remain responsible, as AI is just an assisting tool. 
        
        \item AI assistant usage policies are rare and mainly motivated by privacy concerns, but some participants desire policies to ensure secure usage.
    \end{itemize}
\end{summaryBox}

\subsection{Expected Future Impact of AI Assistants on Security and Development Practices} 
\label{ssec:results-future-impact}

Lastly, our participants shared their views on the future impact of AI assistants on software development, the changes they expect to software development, their influence on software security, and their wishes for future usage of AI tools (RQ3).

\subsubsection{Future Impact on Security} 
Participants were undecided regarding the future impact of AI assistants on security. 
Many expected AI tools to help with security, as they believed AI tools would be able to support developers with security during software development. 
For example, they imagined AI could find common vulnerabilities and take over static code analysis. 
Some stated, however, that the quality and accuracy of the tools would need to improve for use in security tasks: 
\blockquote[P26]{If we can make sure AI spits out perfectly secure code, which it will be capable of doing along the line, then AI will be improving the overall cybersecurity posture.}.

In contrast, about half of the developers expected using AI assistants to impact software security negatively. 
For example, some noted that the quality of AI tools was not yet high enough to ensure security, but they expected this to be the case in the future: 
\blockquote[P27]{For the first time, using AI tools for writing applications, we will have more vulnerabilities, and we will need to fix them. However, with time, the AI model will learn some details, and I think it will be fixed.}.
A few others suspected developers of blindly trusting the AI output during software development and about software security, resulting in vulnerabilities being introduced into the software: 
\blockquote[P9]{If you're just blindly \textelp{using AI and check} the code in without proper testing or without proper reviewing, I think there is a very high chance that there can be a security flaw in that.}.

Further, some had concerns that AI assistants could effectively be used by potential attackers, with AI tools being able to find and exploit common vulnerabilities, making it possible for attackers with little technical knowledge to perform attacks:
\blockquote[P12]{I think decades ago, script kids referred to kids who get access to some dangerous piece of code. They do not necessarily know what the code does, but when they run it, it's very damaging. Nowadays, the kids just need to express what they want to achieve, and then ChatGPT will write some bad code for them.}.

\subsubsection{Expected Changes to the Software Development Process}
Overall, participants expected AI assistants to become more involved in the software development process, taking over more mundane tasks, thus shifting the developers' responsibilities toward complex tasks that AI assistants cannot solve.

Participants mentioned a variety of tasks for which they want to use AI in the future. 
The majority expected to be able to use AI on security-relevant tasks. 
This includes code generation, security reviews, vulnerability detection, software testing, and malware analysis. 
However, many participants agreed that the suggestion quality needs to improve for such tasks. 
Some participants expect this to happen in the future:
\blockquote[P4]{I think over the course of time, in the next three, two, or four years, definitely AI will write better code than developers. That is something more secure and better.}.
Participants expect AI assistance for other tasks, like maintaining code, installing and updating libraries, or software design. 

Some participants speculated that AI tools might improve through highly task-specific training, e.g., tools specifically trained for software security instead of models for general use. 
They expect these tools would perform better and would trust them more: 
\blockquote[P2]{%
It will be really hard to trust and rely on the \textins{general} models when there are security properties required. However, I guess that if a model was developed with a security-first principle, focused on security, the story might be different \textelp{}}.

Many participants expected significant shifts in software developers' responsibilities, such as developers becoming prompt engineers or becoming an AI supervisor who primarily evaluates AI suggestions. 
About half of the participants did not expect AI assistants to replace humans in software development. 
Instead, they suspect AI assistants will speed up tedious and time-consuming tasks, shifting software developers' tasks with more time for high-level tasks. 
Thus, the majority shared a generally positive outlook on AI's influence on its future integration into the \gls{sdlc}:
\blockquote[P16]{I don't think it will ever replace a developer. \textelp{} I think it's more likely that we, as a software engineering industry, would do less and less of those mundane tasks and more of the interesting stuff.}.
However, some were worried about their job, as AI might perform many current software developer tasks if they further improve. 
This fear was mostly caused by AI assistants' ability to solve many tasks very quickly compared to human developers.

\subsubsection{Wishes}
\label{sssec:results-wishes}
Participants hoped that the usability of AI assistants would improve along with the quality of the generated output.
For example, some developers hoped AI tools would be better integrated into IDEs. 
A few others wanted easier methods to prompt the AI, e.g., voice commands or the ability to pass drawings and diagrams to the AI to explain program structures easily. 
This was combined with the wish for technical improvements of the AI systems, for example, a larger context window: \blockquote[P20]{Generative AI will get more powerful and can process more context and broader context to generate better codes, and maybe in the future, a whole project, which will need very minor modifications from the human user or developer, probably.}.

However, many participants mentioned that the output quality needs to improve before AI assistants could be of greater help to them, as they had issues with the quality and correctness of the AI output in the past (\autoref{sssec:results-concerns}). 
About half were confident that the quality would improve in the future, with a few unsure how fast these tools could improve: 
\blockquote[P10]{I think they still have a long way to go. There is a lot more training that they should undergo. They have to improve.}. 
A few participants mentioned observing a drop in the quality of the AI output over time, claiming the AI needs to be trained with higher-quality data. 

As most participants had concerns regarding leaking sensitive data and IP through third-party AI assistants (\autoref{sssec:results-professional-context}), some desire and expect increased use of self-hosted and specialized AI assistants within companies:
\blockquote[P21]{The thing is, if you are having your local model, which is, again, coming to the security and the privacy, that's the only way that your data is not accessed. \textelp{If} this model on your server, it's not going outside your network.}.

\begin{summaryBox}{Key Findings: Expected Future Impact (RQ3)}
    \begin{itemize}[leftmargin=*]
        \item Participants expect their role to shift from writing code to more creative and complex tasks, leaving the mundane for AI assistants under their supervision.
        \item Participants desire improvements in AI assistant quality, correctness, and security abilities.
        \item Some participants envision AI assistants to improve in security tasks, while others argue for a negative impact.
    \end{itemize}
\end{summaryBox}

\section{Discussion}
Below, we discuss our results by setting them in context and deriving recommendations on usage and future AI assistants.

\subsection{(Mis)Trust in AI Assistants}

While our participants largely maintain a critical mistrust towards AI assistants, they widely use them in software development (e.g., to generate code) at the same time. 
Although this mistrust applies to security (e.g., generating vulnerable code), participants reported security issues rarely. 
They used other aspects, like functionality and correctness of AI suggestions, as a proxy to assess AI assistants' security performance. 
That said, mistrusting AI assistant security is likely due to overall quality issues that participants widely experience.
Currently, this skepticism leads participants to use AI assistants with care and scrutinize AI suggestions.
Changes to these proxy indicators might affect developer behavior. 
For example, future quality improvements might lead to blindly trusting AI assistants and not checking code suggestions' security before using them.
Further, getting more used to AI assistants might have similar effects and could be expected for such a novel technology.
However, this hypothesis needs to be investigated in future research.
The Reddit discourse analysis broadly aligns with the interview findings, as Redditors and our participants use AI assistants for various software development tasks. 
We found similar mistrust in AI assistants, as interviewees and Redditors expressed the need to scrutinize AI suggestions.

\subsection{Comparison with Related Work}

\subsubsection{Mismatch with Prior Experimental Results}
\label{sssec:discussion-mismatch}
As this study contributes qualitative insights that complement prior experiments on the security impact of AI assistants, a comparison finds a major mismatch:
While our participants reported to critically scrutinize AI suggestions (\autoref{ssec:results-checking-correctness}) due to general mistrust and did not perceive negative security impact from AI usage, a negative security impact is evident in practice~\cite{snyk-ai-2023, pearce2022asleep, perry2023ai-assistant-security, majdinasab2023assessing, conf/usenix/SandovalPNKGD23}.
This mismatch reveals a skewed self-perception, so that software professionals overestimate their capabilities in scrutinizing AI suggestion for security.
Nonetheless, we conclude that software professionals are aware of potential security issues due to AI assistant usage and try to ``stay awake''~\cite{pearce2022asleep}, but seem to lack methods and support to effectively validate AI suggestions.
\extendedversion{To the literature that focussed AI code generation security, we add novel insights on security use cases like vulnerability detection, threat modelling, and offensive security tasks.}{}

\subsubsection{Security Capabilities of AI Assistants}
Our participants feel that AI capabilities are still limited and unreliable, while being a supportive tool at the same time. 
However, they generally believe \glspl{llm} will become more helpful in assisting with security in the future (\autoref{ssec:results-future-impact}).
Currently, research found mixed capabilities in AI assistants, ranging from poor quality and insecure suggestions~\cite{pearce2022asleep, perry2023ai-assistant-security, majdinasab2023assessing, conf/usenix/SandovalPNKGD23} to autonomously outperforming CTF players~\cite{shao2024empirical, shao2024NYU, project-naptime, moskal2023llmskilledscriptkiddie, tann2023usinglargelanguagemodels, pearce2022popquizlargelanguage}.
As our participants did not perceive such immense benefits, this supports that studies might overestimate AI security performance~\cite{ding2024vulnerabilitydetectioncodelanguage, wu2023exploringlimitschatgptsoftware} when used in practice.
Based on our interviews, we hypothesize that challenges when using AI assistants, e.g., providing enough context, engineering prompts, or ethics safeguards, currently prevent leveraging the full potential that might be achievable in theory and under ideal lab conditions.  

\subsection{AI Assistants as a (Novel) Source of Advice}
\label{ssec:discussion-novel-advice-source}

We found our assumption confirmed that AI assistants are a new advice source. 
Participants reported to have primarily replaced classical online advice sources, like Google and \gls{so}, by using AI assistants (\autoref{sssec:results-tasks}). 
We see similar usage patterns, such as copying, pasting, and adapting code (\autoref{ssec:results-checking-correctness})---that are known to cause security issues, e.g., when copying from \gls{so}~\cite{Fischer:2017}.
Comparably, the research community also found software professionals to achieve worse security when using AI assistants compared to not using them~\cite{pearce2022asleep, perry2023ai-assistant-security, conf/usenix/SandovalPNKGD23, majdinasab2023assessing}.
While recent research found ChatGPT not to entirely replace \gls{so}, 35\% preferred the former due to its language characteristics and comprehensiveness---even with a large portion of incorrect answers~\cite{kabir2024stackoverflowobsolete}.

Considering AI assistants a (partial) replacement for other online advice sources, it remains an open question how AI assistants impact online knowledge communities in which they have been trained (partly). 
Recently, \citeauthor{10.1145/3624732} found AI assistants degrade online communities and reduce the number of users on \gls{so}~\cite{10.1145/3624732}. 
This could cause a ``vicious cycle'' of feedback when it drains the online communities on which it is trained.
Following that argument and given the often poor quality and insecure suggestions on \gls{so}, creators of AI assistants need to be aware of this problem and prevent reinforcing insecure suggestions~\cite{snyk-copilot-amplifies}.

\subsection{Recommendations}
Below, we give recommendations for AI assistant users and creators:

\subsubsection{Critically Scrutinizing AI Suggestions}

We advocate, similar to other work that demonstrated the security shortcomings of AI code assistants~\cite{pearce2022asleep}, to critically validate all AI suggestions. 
Despite the found mismatch that questions its feasibility for software professionals (\autoref{sssec:discussion-mismatch}), we argue that awareness of AI unreliability and potential security issues is important nonetheless---and required for critical scrutiny.
While our participants often already showed this awareness, we underline the need to educate software professionals and companies about potential security issues arising from AI usage, e.g., package hallucinations~\cite{package-hallucinations, package-hallucinations-blog}. 

How to scrutinize AI suggestions (and generally ensuring code security) remains an open question.
As a rule of thumb, we recommend treating AI-generated code like human code and applying the same quality assurance measures, e.g., code reviews, software testing, static analysis, or pentesting.
Many software professionals and companies already had such structured processes (cf.\ \autoref{fig:checking-ai-generated-code}).

Given potential security decreases, we argue that AI assistant usage should be considered depending on the security guarantees needed in a software project on a case-by-case basis.
Currently, our participants are concerned that AI assistants do not outperform humans with expert security knowledge, raising the question of why AI is used for security-critical tasks. 
Given that participants (need to) check the AI suggestions, humans with sufficient knowledge and skills are still required to do these checks anyway.

\subsubsection{Improving Model Quality and Security}
Given the current widespread usage of AI assistants, which can be expected to become even more ubiquitous, reducing security issues at the model level is likely to have the highest impact.
Our participant's quality and security concerns were reflected in the demand for improved future models (\autoref{sssec:results-wishes}). 
As our participants, we anticipate further AI assistant improvements.
Along with the general improvements in AI assistant quality and performance, AI creators should also ensure suggestions are reliable and secure to avoid risk to downstream users of software built with AI suggestions.

We argue that models with security capabilities are needed if AI assistants are used in software engineering.
For example, coding models must be constrained to generate secure code suggestions (at least at a high rate).
As this can largely depend on the \glspl{llm}' training data, we advocate rethinking what data is used for training. 
While current models are trained on large corpora of online content, e.g., from GitHub or \gls{so}, it is no surprise that the resulting AI suggestions might be insecure given many insecure code snippets online~\cite{Fischer:2017, Fischer:2019:StackOverflowHelpful, Fischer:2021:ContentReranking}.
Using datasets with better quality and security could also make AI suggestions more secure. 
For existing models, security hardening techniques should be considered~\cite{he2023large}.

We hypothesize that using task-specific models, e.g., for vulnerability detection, threat modelling, or secure code generation, instead of general-purpose models might result in better quality and security capabilities.
For example, HackerOne recently launched \emph{Hai} beta, an AI assistant specifically tailored to vulnerability intelligence tasks, e.g., to assist with vulnerability remediation~\cite{hackerone-hai}.

\subsubsection{Leveraging Prompt Engineering}

To improve and get the best possible AI assistant suggestions, we recommend software professionals to leverage prompt engineering.
Also, the AI assistant creators suggest prompt engineering, e.g., OpenAI~\cite{openai-prompt-engineering}, indicating this is necessary to circumvent low-quality suggestions like our participants reported.
When not satisfied with the first suggestion, software professionals should try to iteratively refine their prompts, e.g., starting with simple queries, then providing more context, being more specific, or splitting a problem in smaller sub-problems.
Many participants already apply known prompt engineering techniques~\cite{fagbohun2024empiricalcategorizationpromptingtechniques} by adapting their prompts to obtain the desired AI suggestions (\autoref{sssec:results-concerns}).
We recommend learning about and exploring prompt engineering practice guides~\cite{openai-prompt-engineering,fagbohun2024empiricalcategorizationpromptingtechniques, santu2023telergeneraltaxonomyllm}.
Still, prompt engineering remains a significant usability obstacle, limiting AI assistant usefulness~\cite{Liang24}.

\subsubsection{Shifting from Compliance-Driven to Security-Driven Policies}
Interestingly, the companies participants work for seem less concerned about the security of AI-suggested code. 
Instead, they view data usage and privacy aspects (\autoref{sssec:policy-regulations}) as the main motivation behind AI assistant usage policies---although Google did ban AI assistants due to security concerns~\cite{ai-ban-google}. 
Given many participants shared AI assistant security concerns, one explanation for compliance-driven policies is that management or legal teams create them, but not software professionals, as our participants were rarely involved. 
Another explanation is that AI suggestions are evaluated like human code (\autoref{ssec:results-checking-correctness}), not needing a dedicated policy.
Nonetheless, we think data and privacy leakage concerns are important aspects and need to be considered by companies; recently, 
\citeauthor{Niu2023} uncovered that about 8\% of prompts to the GitHub Copilot models result in privacy leaks~\cite{Niu2023}.
Overall, we acknowledge that using AI assistants is a trade-off between security, privacy, cost, efficiency, and liability.
We call companies to remember to consider security in this trade-off.

\subsubsection{Prioritizing Usage of Privacy-Friendly AI Assistants}
A significant participant concern was leaking data and sensitive information when using AI assistants (\autoref{sssec:results-professional-context}), which also results in policies regulating usage (\autoref{sssec:policy-regulations}).
Other researchers also found these privacy concerns among general users of AI assistants, and we can confirm trade-offs between privacy and utility~\cite{zhang2024llmdisclosure} for software professionals.
Consequently, participants desire self-hosted AI assistants, i.e., not involving a third party, or private ones, i.e., hosted by a third party but with privacy guarantees (e.g., no training on prompts).
The latter might be interesting if the hardware is unavailable, e.g., to host models like \gls{llama}~\cite{llama}.
We recommend software professionals and companies to consider these more privacy-friendly variants of AI assistants.
The creators of AI assistants should offer their models either for self-hosting or in a private subscription.
The industry recognizes this need already; for example, GitHub recently launched \emph{Copilot Enterprise}~\cite{copilot-enterprise, copilot-enterprise-blog}.
When also fine-tuning such models (e.g., on a company's code base), this might improve quality and security of AI suggestions.

\subsubsection{Balancing Ethical Concerns and Using AI Assistants for Security}
As participants reported, one limitation to using AI assistants for security tasks are the implemented ethics safeguards.
AI assistants might refuse prompts they deem unethical, e.g., more offensive security tasks like identifying a vulnerability or creating an exploit (\autoref{sssec:results-ethics-constraints}). 
While these ethical considerations are important, they create a dilemma, as vulnerabilities must be found and fixed to improve security. 
For ethical usage, e.g., security evaluations of one's software, this can limit AI assistants' usefulness---for valid use cases that would be done by human security experts otherwise.
Further, participants reported circumventing safeguards with prompt engineering. 
While we advocate the creation of AI assistants for specific security tasks, we believe it is necessary to discuss the ethics first and to implement the respective ethical constraints that cannot be easily circumvented.
This is also necessary as specific security AI assistants like \emph{Hai}~\cite{hackerone-hai} emerge.

\subsection{Outlook \& Future Work}

\subsubsection{General-Purpose \& Code AI Assistants}
\autoref{tab:participant-overview} confirms the wide usage of ChatGPT and other general-purpose AI assistants among software professionals~\cite{2023StackOverflowSurvey}, even more than coding assistants like GitHub Copilot. 
However, we perceived a strong focus on AI code assistants like Copilot in security research~\cite{pearce2022asleep, perry2023ai-assistant-security, conf/usenix/SandovalPNKGD23, majdinasab2023assessing}. 
Future research should close this gap and consider both coding-related and general-purpose AI assistants (which are also used for coding tasks), e.g., comparing the security impact of using both kinds.
Similarly, the creators of general-purpose models should also consider the security impact when their models are capable of and used for software development tasks.

\subsubsection{AI Assistants vs.\ Other Advice Sources}
Considering AI assistants as novel advice sources (\autoref{ssec:discussion-novel-advice-source}), the question of whether software professionals deal differently with AI assistants' suggestions and other advice sources arises.
Hence, we advocate experiments to compare the security impact of AI assistants to other advice sources (e.g., Google, \gls{so}), similar to earlier related work~\cite{Acar:2016ww}. 

\extendedversion{
    \subsubsection{Future AI Advances}
    While our work is a current snapshot of security considerations and implications, AI assistants remain novel and further advances can be expected from the rapidly evolving AI field. \glspl{llm}' performance and quality might improve significantly---including security capabilities.
    Additionally, as novelty wears off and AI assistants become ubiquitous, scrutiny through critical mistrust might fade among its users. 
    A future replication of this study could investigate how practices and perception of AI assistants and security changes.
}{}

\section{Conclusion}

We investigated software professionals' usage of AI assistants in software development, focusing on security and their security considerations in 27~interviews, complemented by the analysis of Reddit posts.
Besides being used often by almost all participants, we found that both coding AI assistants, like Copilot, and general-purpose AI assistants, like ChatGPT, are widely used for security-critical software development tasks (e.g., code generation, threat modeling, code reviews, and vulnerability detection).
Despite ubiquitous usage, we found that our participants mistrust and check AI suggestions. 
While security is a primary concern, only a few participants reported negative experiences with AI assistants in the past. 
As our results qualitatively complement prior experiments~\cite{pearce2022asleep, perry2023ai-assistant-security, majdinasab2023assessing, conf/usenix/SandovalPNKGD23}, a comparison reveals a mismatch between our participants' reported scrutiny and actual code security when using AI assistance.
This indicates that software professionals overestimate how well they can scrutinize AI suggestions.
A contributing factor is likely that participants reasoned about AI assistant security capabilities based on proxies such as functionality.
Overall, we conclude that AI assistants change software development by being a novel source of security and security-relevant advice for software professionals.

\begin{acks}
   We thank our anonymous reviewers and shepherd for their valuable feedback and for helping us to improve this paper. 
We also acknowledge \emph{Dagstuhl Seminar 23181} in which most authors participated and where this project started.
This research was funded by \grantsponsor{vws}{VolkswagenStiftung}{https://www.volkswagenstiftung.de/en} \grantnum{vws}{Niedersächsisches Vorab – ZN3695}.
This research was also partially funded by the \grantsponsor{dfg}{Deutsche Forschungsgemeinschaft (DFG, German Research Foundation)}{https://www.dfg.de/en/index.jsp} under \grantnum{dfg}{Germany's Excellence Strategy -- EXC 2092 \textsc{CaSa} -- 390781972} and 
supported by the \grantsponsor{nsf}{U.S.\ National Science Foundation (NSF)}{https://www.nsf.gov/} \grantnum{nsf}{Award \#~2247141} and \grantnum{nsf}{Award \#~2312321}. 
The research was also partly supported by \grantsponsor{EU}{European Union Horizon Europe program - Cybersecurity}{https://cordis.europa.eu/programme/id/HORIZON.2.3.3/en} \grantnum{EU}{\textit{Sec4AI4Sec} Award \#~101120393} and by \grantsponsor{NWO}{NWO, Dutch Research Organization - Kennis- en Innovatieconvenant (KIC)}{https://www.nwo.nl} \grantnum{NWO}{\textit{HEWSTI} Award \#~KICH1.VE01.20.004}.
This work was also supported by \grantsponsor{epsrc}{EPSRC}{https://www.ukri.org/councils/epsrc/} Grants
\grantnum{epsrc}{\textit{REPHRAIN: National Research Centre on Privacy, Harm Reduction and Adversarial Influence Online} (EPSRC Grant EP/V011189/1)} and \grantnum{epsrc}{\textit{Equitable Privacy} (EPSRC Grant EP/W025361/1)}. 

\end{acks}

\section*{Availability}
\label{sec:availability}

To support transparency, replication, and meta-studies, we provide the following research artifacts:
(1)~our recruitment materials,
(2)~the screening questionnaire,
(3)~the interview guide,
(4)~the demographics questionnaire,
and 
(5)~the list of posts from the Reddit analysis.
We do not provide interview transcripts to protect our participants' privacy.
The replication package is available as supplementary material at  
\url{https://doi.org/10.17605/OSF.IO/XZ72H}.

\bibliographystyle{ACM-Reference-Format}
\extendedversion{}{\balance}
\bibliography{bibliography}

\appendix
\extendedversion{\section{Background: AI Assistants \& LLMs}
\label{app:background}

AI assistants like GitHub Copilot and ChatGPT are powered by \glspl{llm} such as OpenAI's Codex and GPT-3, respectively~\cite{GitHubCopilot, OpenAICodex}. 
OpenAI's 2020 release of GPT-3, a neural network based \gls{llm} trained on 45~TB of text, served as a foundation for OpenAI's Codex model, which was additionally trained on 159~GB of Python code and 54~million GitHub repositories~\cite{cooper2023gpt3}. 
Codex aims to generate code based on natural language inputs. 
The results generated by the \gls{llm} are presented to the user in the user interfaces of AI assistants such as GitHub Copilot and OpenAI's ChatGPT. 
ChatGPT's \glspl{llm} were not designed explicitly for code generation, although its training dataset also contained code snippets. 
Since then, several general and coding AI assistants, such as Google's Bard/Gemini, Amazon's Code Whisperer, and many more, have been released. 

Besides suggesting code snippets, the scope of AI assistants has spread to other software development tasks. 
Amazon's Code Whisperer, for example, advertises its ability to perform code reviews, the mechanism for which consists of processes such as \emph{security scan}, \emph{reference tracker}, and \emph{bias avoidance}~\cite{dilmegani2023aiusecases, awscodewhisperer}. 
Other tasks include requirement analysis, code completion to help save the developers' time on tedious tasks, and the automation of code refactoring to improve code maintainability~\cite{Liang24}.

\citeauthor{bird2022taking} offer early insights and opportunities of AI-powered pair-programming tools~\cite{bird2022taking}. 
They describe the pair-programming process by comparing the coder-reviewer relationship with that of a driver and navigator. They find that the developer who originally wrote the code becomes a developer navigator, guiding the AI assistant while directing and reviewing the generated code. 
The study shows that GitHub Copilot can generate code, such as smaller functions that fit the rest of the code base or learn new programming languages. 
However, users also face challenges, like potentially leaking code secrets (e.g., API tokens) and quality issues, such as missing checks for negative array indices.
\citeauthor{ernst2022ai} study AI-driven development environments (AIDEs)~\cite{ernst2022ai}. 
The authors discuss the benefits of automation of mundane tasks and \gls{api} interactions, and the idea that novice developers can learn how to code with the AI assistant's help.
While the papers above focus on general usage practices of AI assistants, they do not focus on security implications and applications in software engineering.
We fill this gap and qualitatively explore how software professionals use AI assistants and their security considerations. 
}{}
\extendedversion{\section{Interview Guide}
\label{app:interview-guide}

For the semi-structured interviews, we followed the below interview guide.

\subsection{Intro}\label{intro}

\begin{itemize}
\item
  \textbf{Thanks:} Welcome to our interview, and thank you very much for
  offering your valuable time for this interview. {[}If not done before,
  introduce yourself.{]}
\item
  \textbf{Ready:} Do you have any initial questions, before I introduce
  you to the interview?
\item
  \textbf{Structure:} First off, I am going to give you a brief
  introduction to the interview topic, the overall procedure, and data
  protection. If you are okay with everything and have no further
  questions, only then we would start with the actual interview.
\end{itemize}

\subsubsection{Context}\label{context}

\begin{itemize}
\item
  \textbf{Who We Are:} We are researchers from multiple universities and
  research institutes that collaborate for this project.
\item
  \textbf{Our Research:} In this project, we are conducting research on
  security aspects of using AI assistant tools in software development,
  like GitHub Copilot or OpenAI's ChatGPT and other large language
  models (LLMs). This boils down to "How are AI assistants used in
  software development and what are the associated security
  implications?".
\item
  \textbf{This Interview:} This interview is intended as an exploration
  of practitioners\textquotesingle{} experiences using AI assistant
  tools and related security aspects, like challenges to obtain secure
  code, vulnerabilities introduced through AI, or improvements in code
  security.
\item
  \textbf{For this Interview:}

  \begin{itemize}
  \item
    We are interested in your \emph{personal opinions and experiences}.
  \item
    We are \emph{not judging} any practices, projects, nor their
    security.
  \item
    If you don\textquotesingle t know the answer, can\textquotesingle t
    remember something, don\textquotesingle t want to answer, or are not
    allowed to answer a question, feel free to just say "next" and we
    will skip the question.
  \end{itemize}
\item
  \textbf{Questions?} Do you have any questions about the interview
  context so far?
\end{itemize}

\subsubsection{Procedure}\label{procedure}

\begin{itemize}
\item
  \textbf{Duration:} The interview duration depends a bit on your
  answers, in our experience interviews last between 45-50 min.
\item
  \textbf{De-Identification:} As outlined in the consent form, we will
  fully de-identify you in any publication and only include short
  quotes.
\item
  \textbf{Recording:} We would like to record this interview so that we
  can transcribe the answers later. The recording will be destroyed
  after we transcribe your answers (a few days).
\item
  \textbf{Questions?} Do you have any questions about the procedure?
\item
  \textbf{{[}Turn recording on{]}} → "The recording is now on."
\item
  \textbf{Restate:} Again for the recording: "Are you okay with this
  interview being recorded?"
\end{itemize}

\subsubsection{Warm Up}\label{warm-up}

\begin{itemize}
\item
  \textbf{Introduction:} Can you briefly introduce yourself and what you
  are working on in software development?

  \begin{itemize}
  \item
    \textbf{Project:} Can you briefly summarize what kind of projects
    you are working on?
  \item
    \textbf{Experience:} How long have you been involved in software
    development?
  \item
    \textbf{Role:} What is your current role?
  \end{itemize}
\end{itemize}

\subsection{Use of Generative AI
{[}RQ1{]}}\label{use-of-generative-ai-rq1}

As already mentioned, we are interested in AI assistants in the context
of software development. So we would like to learn from your
experiences.

\begin{itemize}
\item
  \textbf{Tools:} Can you give us an overview of which AI assistants
  tools, e.g., GitHub Copilot, OpenAI's ChatGPT, or others you have
  used?

  \begin{itemize}
  \item
    \textbf{Other Tools:} Have you used any other AI assistants tools?
  \item
    \textbf{Generative AI Experience:} How long have you been using AI
    assistants tools?
  \item
    \textbf{Getting In Touch:} How did you first learn about AI
    assistants and started using them?

    \begin{itemize}
    \item
      Who introduced you to AI assistants? {[}Prompt for: themselves,
      the organization, colleagues, etc.{]}
    \end{itemize}
  \item
    \textbf{Motivation:} What was your motivation to use AI assistants
    tools?
  \end{itemize}
\item
  \textbf{Tasks:} What are you using generative AI for in your
  professional life?

  \begin{itemize}
  \item
    {[}Specifically prompt for: generating/writing code, fixing code
    that does not work, explaining code, reviewing code for security
    problems and fixing them{]}
  \item
    \textbf{Workflow:} Can you roughly describe the workflow of how you
    use AI assistants for these tasks?

    \begin{itemize}
    \item
      How do you typically deal with the generated output?
    \item
      To what extent do you check the output to be correct, e.g., that
      code works as intended? Why?
    \item
      To what extent do you check the security, e.g., that generated
      code has no vulnerabilities? Why?

      \begin{itemize}
      \item
        How do you check the generated code?
      \end{itemize}
    \item
      To what extent do you change the output of the AI assistants
      before using it?

      \begin{itemize}
      \item
        Did you have to change AI output because it was insecure, e.g.,
        a vulnerability in generated code?
      \item
        {[}If yes: What did you need to change?{]}
      \end{itemize}
    \item
      How much do you usually tweak your prompts for the AI assistants?
      Why?
    \end{itemize}
  \item
    \textbf{Usage:} How often do you use AI assistants?

    \begin{itemize}
    \item
      {[}Prompt for: smaller tasks, internal/external software{]}
    \end{itemize}
  \item
    \textbf{Usefulness:} In what situations do you think AI assistants
    are useful? Why?
  \item
    \textbf{Un-Usefulness:} In what situations do you think AI
    assistants are not useful? Why?
  \item
    \textbf{Usage over Time:} In retrospect, would you say that you
    changed how you use AI assistants since you started using them? How?
  \end{itemize}
\item
  \textbf{Policies:} Does your work have any policies for the use of AI
  assistants?

  \begin{itemize}
  \item
    \emph{{[}If covered:{]}}

    \begin{itemize}
    \item
      \textbf{Allowed to Use Generative AI:} Is it allowed to use AI
      assistants?

      \begin{itemize}
      \item
        \emph{{[}If not:{]}} Why not?
      \item
        For which tasks or in which scenarios?
      \item
        When is it not allowed?
      \item
        Are there any requirements that relate to security?
      \end{itemize}
    \item
      \textbf{Policy Changes:} Did or does using AI assistants require
      any changes to these policies?

      \begin{itemize}
      \item
        \emph{{[}If any:{]}} Can you briefly describe what changes were
        made?
      \end{itemize}
    \item
      \textbf{Shadow Practices:} Besides any official practices, are
      there any undocumented or shadow practices?

      \begin{itemize}
      \item
        \emph{{[}If question unclear, give example: Still using ChatGPT
        or Copilot, although it is forbidden, or blindly accepting code
        suggestions of ChatGPT.{]}}
      \end{itemize}
    \end{itemize}
  \item
    \emph{{[}If not covered:{]}}

    \begin{itemize}
    \item
      \textbf{Inclusion in Policies:} Would you like to have such a
      policy? Why?/Why not?
    \item
      \textbf{Most Important Policy:} What do you think would be the
      most important policy to add?
    \end{itemize}
  \end{itemize}
\end{itemize}

\subsection{Security Implications of Generative AI
{[}RQ2{]}}\label{security-implications-of-generative-ai-rq2}

Thank you for sharing your experiences. Now I want to shift the focus
towards security implications and aspects of AI assistants.

\begin{itemize}
\item
  \textbf{Security Impression:} What is your impression on the security
  of code that is produced by AI assistants?
\item
  Next, I would like to talk about advantages and disadvantages of using
  AI assistants when developing secure software.
\item
  \textbf{Advantages:} What security benefits and improvements do you
  think AI assistants have (if any)?
\item
  \textbf{Disadvantages:} Asking the other way around, what security
  challenges and disadvantages do you think AI assistants have (if any)?

  \begin{itemize}
  \item
    \textbf{Liability Challenges:} What do you think about liability
    when using code that was produced by AI assistants?

    \begin{itemize}
    \item
      Is it necessary to review the generated code? {[}If yes: Who
      approves the code written by AI?{]}
    \item
      Is there an approval process defined specifically for AI-generated
      code or does it go through the same process as other developers in
      the team?
    \item
      Who do you think is responsible when some security vulnerability
      was introduced by AI assistants?
    \item
      Who do you consult if question arise about AI assistant
      suggestions?
    \end{itemize}
  \end{itemize}
\item
  \textbf{Examples:} Do you remember any situations in which AI
  assistants improved or worsened the security of your software? Can you
  tell us about that?
\end{itemize}

\emph{We are now done with more than two third of the interview.}

\subsection{Outlook {[}RQ3{]}}\label{outlook-rq3}

Alright, thanks for these insights. In the last block of questions I
would like to talk about the general future of secure software
development in the context of AI assistants.

\begin{itemize}
\item
  \textbf{Past Changes:} From your perspective, how did using AI
  assistants change creating secure software overall?
\item
  \textbf{Overall Security Impact:} What do you think is the overall
  security impact of AI assistants?
\item
  \textbf{Future Changes:} What do you expect, how will AI assistants
  impact software security in the future?

  \begin{itemize}
  \item
    \textbf{Reshape:} Do you expect AI assistants' impact to reshape
    software security in parts or entirely? Why? {[}Could prompt for
    very hypothetical things: e.g., classes of vulnerabilities becoming
    irrelevant (e.g., SQL injections are caught by AI assistants), AI
    assistant's code understanding is so good that it spots
    vulnerabilities, so we might not need techniques like fuzzing{]}
  \item
    \textbf{Pace:} When do you expect the changes you mentioned to
    happen?
  \item
    \textbf{Impression:} Do you think that AI assistants make it easier
    in the future to create secure software? Why?/Why not? {[}Could
    prompt for: faster, cheaper, requiring less knowledge, etc.{]}
  \item
    \textbf{Vulnerabilities:} Do you expect that AI assistants will
    reduce or increase security issues like vulnerabilities? Why?
  \item
    \textbf{Outlook:} What tasks related to software security could you
    imagine will AI assistants be able to do in the future?
  \end{itemize}
\item
  \textbf{Process Impact:} How do you expect AI assistant to impact
  processes for secure software development? {[}Could prompt for :
  Secure Software Development Lifecycle (SDLC), DevSecOps, \ldots{]}

  \begin{itemize}
  \item
    \textbf{Job Impact:} How do you expect AI assistants to change your
    daily work?

    \begin{itemize}
    \item
      (e.g., less writing code (done by AI), but more reviewing its
      security -- or vice versa?)
    \item
      To what extent might AI assistants take over some of your duties
      at work?
    \item
      What do you expect AI assistants will not be able to do that you
      can?
    \end{itemize}
  \end{itemize}
\item
  \textbf{Developer vs.~Generative AI:} I know this question is tricky,
  but maybe you have an opinion on this: Who do you think might be
  producing more secure code -- you or AI assistants? Why?
\item
  \textbf{Needs/Wishes/Changes:} Imagining there are no limitations,
  what future AI assistants tools and functionality would you like to
  have or make to current AI assistants for improved secure software
  development?
\end{itemize}

\subsection{Outro}\label{outro}

Alright, thank you so much for this helpful insights! Those were almost
all questions that we had for you. Finally, there are only two questions
left:

\begin{itemize}
\item
  \textbf{Missed Questions/Topics:} Are there any topics or questions
  that you expected me to ask about, but I didn\textquotesingle t?
\item
  \textbf{Anything else (recorded):} Do you have any final questions or
  comments you want to make on the recording? There will be time after I
  stopped the recording, so you can ask me without the recording if this
  is more comfortable for you.
\item
  {[}Stop Recording{]} The recording is now off.
\item
  \textbf{Anything else (not recorded):} Do you have any final comments
  or questions?
\end{itemize}

\begin{itemize}
\item
  {[}Offer Compensation → collect e-mail address{]}
\item
  {[}Ask for contacts that might be worth interviewing{]}
\end{itemize}

}{}
\extendedversion{\section{Codebook}
\label{app:codebook}

The below codebook contains numbers on code occurrences measured by number of interview transcripts in which a code was assigned, i.e., counting multiple codings of the same code in one interview as only one occurrence.

\noindent\textbf{Demographics:}
    \begin{itemize}

        \item \textit{Tool Experience (13)}
        \item \textit{Participant Background (14)}: Team Lead (5), Software Developer (13), Security (8), Machine learning engineer (4), Tech Lead (1)
    \end{itemize}

\noindent\textbf{AI Tools:}
    \begin{itemize}

        \item \textit{ChatGPT (27)}: Azure OpenAI (1)
        \item \textit{GitHubCopilot (13)}
        \item \textit{Google AI (1)}: Bard (8), Gemini (3)
        \item \textit{Self-hosted/internal AI assistants (6)}
        \item \textit{Bing Chat (5)}
        \item \textit{LLaMA (4)}
        \item \textit{Anthropic Claude (2)}
        \item \textit{Microsoft Copilot (2)}
        \item \textit{Pre-Trained models (2)}
        \item \textit{Amazon CodeWhisperer (1)}
        \item \textit{Tabnine (1)}
        \item \textit{CodeT5 (1)}
        \item \textit{UniXcoder (1)}
        \item \textit{Colab AI (1)}
        \item \textit{Jarvis (1)}
        \item \textit{Jurassic-1 (1)}
        \item \textit{JupiterOne J1 AI (1)}
        \item \textit{Hugging Face Agents (1)}
    \end{itemize}

\noindent\textbf{AI Tool Usage:}
    \begin{itemize}

        \item \textit{Start of usage (21)}
        \item \textit{AI usage over time (14)}
        \item \textit{For what it is used (0)}: Larger things/tasks (1), Small things/tasks (4), External (1), Internal (1)
        \item \textit{Workflow (22)}
        \item \textit{Internal AI Frontend (2)}
        \item \textit{Security Relevance of AI Code (6)}
        \item \textit{Personal Usage (2)}
        \item \textit{Usage Frequency (1)}: Daily (8)
    \end{itemize}

\noindent\textbf{Motivation for using AI tools:}
    \begin{itemize}

        \item \textit{Stay competitive (4)}
        \item \textit{Interest in AI (1)}: Curiosity in AI tools (12), Curious to test and manipulate ChatGPT (0), Benefits of ChatGPT Plugins (1)
        \item \textit{Aid as a pair programmer (3)}
        \item \textit{Generation of Boilerplate Code (4)}
        \item \textit{Uses AI tools for Self-Enhancement (Become a better coder) (6)}
        \item \textit{AI Assistants are proficient with many (progamming) languages (5)}
        \item \textit{Aid for beginners (7)}
        \item \textit{Aid as a security expert (6)}
        \item \textit{Saving money (3)}: AI tools are cheaper than security exclusive tools (1)
        \item \textit{Increase productivity and save time (22)}
        \item \textit{Ask for help if solution is not found immediately (1)}
    \end{itemize}

\noindent\textbf{Getting in touch with AI assistants:}
    \begin{itemize}

        \item \textit{Betas (2)}
        \item \textit{Release of ChatGPT (4)}
        \item \textit{News \& Public Media (4)}
        \item \textit{Social Media (5)}
        \item \textit{Students/Colleagues/Friends (3)}
        \item \textit{Research Conferences (1)}
        \item \textit{Tech Blogs (1)}
        \item \textit{Clients requested tool integration in products (1)}
    \end{itemize}

\noindent\textbf{Task:}
    \begin{itemize}

        \item \textit{Tasks they don't use AI for (0)}: Traffic analysis (1), Code reviews (1), Software Design (2), Bug fix (1), Discover vulnerabilities (7), Tasks concerning confidential data (3), User interfaces (1), Scientific Programming (1), Application design (1), Threat modeling (1)
        \item \textit{Tasks they use AI for (0)}: Security (12), Coding (27), Search and look up information (17), Natural Language Text Tasks (19), Image Generation (2), Game development (1), LLMs used for finance modelling (1)
    \end{itemize}

\noindent\textbf{Problems with AI systems:}
    \begin{itemize}

        \item \textit{Organizational Problems (0)}: Privacy (21), Costs (9), AI might produce copyright-violating results (3), Require trainig for what is allowed to do with AI (1), Access to AI might be taken away (1)
        \item \textit{Human Problems (0)}: Usability (24), Mistrust (e.g., security, correctness) (22), AI always confident (6), Developers might trust output blindly, but don't understand it (9)
        \item \textit{Technical Problems (0)}: Quality (26), Security (24)
    \end{itemize}

\noindent\textbf{Checking AI-generated code (e.g., security, correctness):}
    \begin{itemize}

        \item \textit{Peer Review (10)}: Security team/expert (4), QA Team (4), Team lead (2)
        \item \textit{Implementing solution yourself (3)}
        \item \textit{Participant researches information and proof checks claims (1)}
        \item \textit{Consulting team lead on questions regarding AI suggestions (2)}
        \item \textit{Manual Looking (24)}: Verfify through github (1), Verify trough google search/forums... (4)
        \item \textit{Software testing (5)}: Just running the code, instead of formal testing (+) (5), Fuzzing tools (1), Static Analysis tools (3)
        \item \textit{Tested like human code/not different for AI (7)}
        \item \textit{Check security (5)}
        \item \textit{Not checking security of AI-generated code (1)}
        \item \textit{Checking code with AI assistants (0)}: Ask AI assistant to verify its suggestions/proof claims (2), Ask AI model to write proofs for code verification (1), Verifying other AI answers (4)
    \end{itemize}

\noindent\textbf{Responsibility:}
    \begin{itemize}

        \item \textit{Human remains responsible (25)}
        \item \textit{Responsibility needs to be shifted to AI author (4)}
        \item \textit{Company that publishes code (2)}
        \item \textit{Neither (2)}
        \item \textit{Future policy will decide responsibility (2)}
    \end{itemize}

\noindent\textbf{Human vs AI:}
    \begin{itemize}

        \item \textit{Human (14)}
        \item \textit{AI (7)}
        \item \textit{Combination of both, human and AI (5)}
        \item \textit{Neither (1)}
        \item \textit{AI on level with good junior developer (3)}
        \item \textit{AI will become better than humans (3)}
    \end{itemize}

\noindent\textbf{Policy/Regulation:}
    \begin{itemize}

        \item \textit{Company Policy (0)}: AI usage forbidden/allowed (25), Policy Opinion (16), Shadow Practices (11), Use of local AI APIs to address privacy concerns (1)
        \item \textit{AI regulations not helpful (4)}
        \item \textit{Questions regarding fairness and AI (1)}
        \item \textit{Law playing catch-up (1)}
    \end{itemize}

\noindent\textbf{Changes to secure software development:}
    \begin{itemize}

        \item \textit{Introducing Vulnerabilities (2)}
        \item \textit{People become lazy and depend on AI assistant (8)}
        \item \textit{None (6)}: No negative security impact (4)
        \item \textit{Enablement Sessions for AI usage (2)}
        \item \textit{AI makes development process easier (7)}: AI makes it easier to write small security scripts (1)
        \item \textit{Unsue about speedup (1)}
        \item \textit{Speedup (4)}
        \item \textit{Developers need to be able to assess security of AI code (2)}
        \item \textit{Developers will need to be able to prompt AI (2)}
        \item \textit{Change in approaching new concepts (1)}
    \end{itemize}

\noindent\textbf{Future of AI usage:}
    \begin{itemize}

        \item \textit{AI will create vulnerabilities in the future (1)}
        \item \textit{Wish to remove prompt check/barriers (2)}
        \item \textit{No big changes (3)}
        \item \textit{Trust in AI will improve (1)}
        \item \textit{Need for privacy-friendly/secure LLM services (3)}
        \item \textit{Idea of self-hosted AI tools to prevent data leakage (6)}
        \item \textit{Replacing humans (0)}: AI can replace human (7), AI cannot replace human (14), AI should not replace humans (1)
        \item \textit{Tasks they would use AI for (0)}: IDS/IPS (3), Malware Analysis (1), Threat analysis (1), Automatically Install Library Updates (2), Planing a software project/design (2), Suggestions how to fix vulnerabilities (4), Assisting with simple tasks (3), Acting as a Security Experts (3), Security review (6), Vulnerbility detection (8), Explain code functionallity (1), Code review (4), Fixing Errors (3), Create Demo Code (1), Acting as a coding assistant (3), AI could suggest best frameworks for frontend and backend (1), Deploying code (1), Software testing (2), Refactoring (2)
        \item \textit{Pace (0)}: Unsure (2), Fast (6), Slow (3)
        \item \textit{AI Improvements (0)}: Open source AI tools will be unrestricted (1), Small development companies benefit most from AI (1), Generate code from pictures/drawings (2), Having ensemble of AI assistants that work like a human team (2), AI able to reason on Hardware level (1), AI able to reason about complex systems (2), AI should/will be able to fix bugs (2), Better protect from unethical AI usage (4), AI needs to be trained with higher quality training data (2), Models need to improve (5), Expecting further AI improvments (7), Models should explain their decisions/give reasoning (2), Wants to fine-tune models (for security) (2), Models should indicate answer confidence (1)
        \item \textit{Future Security Impact (0)}: People relying on AI tools makes code less secure (4), Using AI would make all users vulnerable at the same time (1), AI tools can take security responsibilities off developers (7), AI might become malicious (2), Neutral/no impact (3), Helpful to increase security if quality/accuracy gets better (4), AI can identify/fix security problems (6), Impact is hard to measure (7), (Short term) more vulnerabilities (4), Attacker might use AI effectively (6), Some things harder, some things easier (3), Static code analysis by AI (5), No increase in vulnerabilities (3), AI will assist developers and improve security (1), Security analysis by AI (1)
        \item \textit{Security improvements (1)}: Had no vulnerabilities in AI generated code (4), No impact (8), Little improvement on security (2), Overall security improvements (9), Security-focused AI system (5), Unsure (4), Experienced a vulnerability (1)
        \item \textit{Security Testing (0)}: Could find common vulnerabilities (9), Formal Verification (1), Improve fuzzing speed (1), AI will replace static analysis tools (3)
        \item \textit{Usability (0)}: AI to encapsulate the services of the current variety of tools (2), Prompting (4), Developers learning to handle AI tools (4), Better integrated AI tools (e.g., in IDEs) (8)
        \item \textit{Development Process (0)}: More companies will adopt due to ease of automation (1), AI should handle security entirely (2), Focus on non-security related tasks (1), AI assistants can do boring/mundane/repetitive tasks (12), People still need to review AI code (8), Programming code becomes natural language (2), Developer will become an AI supervisor (5), Programmers replaced with prompt engineers (2), AI will take some responsibility off the developer (3), AI will change the process entirely (1)
        \item \textit{Malware \& Ransomware Amplified (1)}
        \item \textit{Generate proofs for code verification (1)}
        \item \textit{Mass-AI automation in future (2)}
        \item \textit{AI models used for security auditing old codebases in companies (1)}
    \end{itemize}
}{}
\extendedversion{\section{Additional Tables and Figures}
\label{app:othertables}
We conclude with tables and figures excluded from the main text for brevity, but included here to support replication. This includes our list of queries used for our preliminary gray literature review and Reddit search (\autoref{tab:searchterms}), and list of categories on usage and concerns around AI-generated code (\autoref{tab:usage_concerns}).

\begin{table*}[ht]
    \centering
    \footnotesize
    \caption{Search queries used in the preliminary gray literature search and Reddit reviews.}
    \label{tab:searchterms}
    \begin{tabular}{l}
        \toprule
            \textbf{Preliminary Round} \\
        \midrule
            ai chatbot writing code \\
            developers opinions on ai chatbot writing code \\
            what are the uses of ai chatbots in writing code \\
            how do developers use ai chatbots in their work \\
        \midrule
            \textbf{Reddit Round 1} \\
        \midrule
            ai chatbot writing code \\
            developers opinions on ai chatbot writing code \\
            how do developers use ai chatbots in their work \\
            what are the uses of ai chatbots in writing code \\
            company policy on ai chatbots \\
            best practices for using ai assistants in coding \\
            ai chatbot debugging \\ 
            best ai assistants to use \\
            ai uses for reviewing code \\ 
            ai uses for explaining code \\
            ai chatbot insecure code \\ 
            risks of ai generated code \\ 
            vulnerabilities of ai generated code \\ 
        \midrule
            \textbf{Reddit Round 2} \\
        \midrule
            ai write code \\
            ai debugging code \\
            ai explaining code \\
            ai documenting code \\
            developers using ai assistant \\
        \bottomrule
    \end{tabular}
\end{table*}

\begin{table*}[ht]
    \centering
    \footnotesize
    \caption{Overview of people's usage and concerns regarding AI-generated code.}
    \label{tab:usage_concerns}
    \begin{tabular}{lllll}
        \toprule
        \textbf{General category} & \textbf{Sub category} & \textbf{\#Posts} & \textbf{\#Comments} & \textbf{Description of code} \\
        \midrule
        What are people using AI generated code for? & Code optimization & 2 & 4 & Using AI to generate efficient code for a given problem. \\
         & Debugging & 2 & 0 & Using AI tools to debug code. \\
         & Translate language & 1 & 0 & Using AI to switch code languages. \\
         & App creation & 1 & 0 & Using AI to develop applications. \\
         & Writes code & 14 & 15 & Using AI to write code. \\
         & Explain code & 4 & 4 & Using AI to explain or document code. \\
        \midrule
        Want to use AI in code & [Want] - Code optimization & 1 & 3 & Desire to use AI for code optimization. \\
         & [Want] - Debugging & 3 & 0 & Desire to use AI for debugging. \\
         & [Want] - Translate language & 0 & 0 & Desire to use AI for code translation. \\
         & [Want] - App creation & 0 & 0 & Desire to use AI for app creation. \\
         & [Want] - Writes code & 13 & 3 & Desire to use AI for writing code. \\
         & [Want] - Explain code & 2 & 0 & Desire to use AI to explain code. \\
        \midrule
        What are people's concerns about AI generated code? & Job insecurity & 9 & 33 & Concerns about job loss due to AI. \\
         & False confidence & 1 & 3 & Overconfidence due to AI use. \\
         & False sense of security & 2 & 1 & Overconfidence in code security due to AI. \\
         & Transparency & 2 & 3 & Concerns about AI companies hiding data. \\
         & Bad code & 2 & 16 & Belief that AI generates poor code. \\
         & Make easier & 8 & 6 & Belief that AI will make programming easier. \\
         & Legal issues & 0 & 1 & Legal concerns with AI in coding. \\
         & Company use & 1 & 1 & Companies using AI for coding. \\
         & Company don't use & 0 & 1 & Companies avoiding AI for coding. \\
        \bottomrule
    \end{tabular}
\end{table*}

}{}

\end{document}